\documentclass[12pt]{article}
\usepackage{authblk}
\usepackage[a4paper,bindingoffset=0.2in,left=0.6in,right=1in,top=0.85in,bottom=1in,footskip=.25in]{geometry}
\usepackage[dvipsnames]{xcolor}
\usepackage{hologo}
\usepackage[]{colortbl}
\usepackage{multirow}
\usepackage[natbibapa]{apacite}

\usepackage{graphics}
\usepackage{enumitem}
\usepackage{graphicx}
\usepackage[]{colortbl}%
\usepackage{soul}
\usepackage{multirow}
\usepackage{amsmath}

\usepackage[english]{babel}
\addto{\captionsenglish}{%
    }

\usepackage[breaklinks,colorlinks = true,
            linkcolor = BlueViolet,
            urlcolor  = BlueViolet,
            citecolor = Aquamarine,
            anchorcolor = white,
            breaklinks]{hyperref}
\usepackage[]{doi}

\def\mybibdoicolor{\color{BlueViolet}} 

\title{{\bf Estimating satellite orbital drag during historical magnetic superstorms\footnote{Paper published in {\it SpaceWeather}, doi: \url{hhttps://doi.org/10.1029/2020SW002472}}}}

\date{\vspace{-6ex}}

\author[1,2\footnote{Electronic address: denny.m.deoliveira@nasa.gov; denny@umbc.edu}]{{\normalsize Denny M. Oliveira}}
\author[2]{{\normalsize Eftyhia Zesta}}
\author[3,4,5]{\normalsize Hisashi Hayakawa}
\author[6,1]{\normalsize Ankush Bhaskar}

\affil[1]{{\footnotesize Goddard Planetary Heliophysics Institute, University of Maryland, Baltimore County, Baltimore, MD, United States}}
\affil[2]{{\footnotesize NASA Goddard Space Flight Center, Greenbelt, MD, United States}}
\affil[3]{{\footnotesize Institute for Space-Earth Environmental Research, Nagoya University, Nagoya, Japan}}
\affil[4]{{\footnotesize Institute for Advanced Researches, Nagoya University, Nagoya, 4648601, Japan}}
\affil[5]{{\footnotesize Rutherford Appleton Laboratory, Chilton, United Kingdom}}
\affil[6]{{\footnotesize Catholic University of America, Washington D.C., United States}}

\begin{document}

	\maketitle

    \begin{abstract}

      Understanding extreme space weather events is of paramount importance in efforts to protect technological systems in space and on the ground. Particularly in the thermosphere, the subsequent extreme magnetic storms can pose serious threats to low-Earth orbit (LEO) spacecraft by intensifying errors in orbit predictions. Extreme magnetic storms (minimum Dst $\leq$ --250 nT) are extremely rare: only 7 events occurred during the era of spacecraft with high-level accelerometers such as CHAMP (CHAllenge Mini-satellite Payload) and GRACE (Gravity Recovery And Climate experiment), and none with minimum Dst $\leq$ --500 nT, here termed magnetic superstorms. Therefore, current knowledge of thermospheric mass density response to superstorms is very limited. Thus, in order to advance this knowledge, four known magnetic superstorms in history, i.e., events occurring before CHAMP’s and GRACE’s commission times, with complete datasets, are used to empirically estimate density enhancements and subsequent orbital drag. The November 2003 magnetic storm (minimum Dst = --422 nT), the most extreme event observed by both satellites, is used as the benchmark event. Results show that, as expected, orbital degradation is more severe for the most intense storms. Additionally, results clearly point out that the time duration of the storm is strongly associated with storm-time orbital drag effects, being as important as or even more important than storm intensity itself. The most extreme storm-time decays during CHAMP/GRACE-like sample satellite orbits estimated for the March 1989 magnetic superstorm show that long-lasting superstorms can have highly detrimental consequences for the orbital dynamics of satellites in LEO.

    \end{abstract}

    {\bf Key Points:}

    \begin{itemize}

      \item Satellite orbital drag during magnetic superstorms (standard/equivalent Dst $\leq$ --500 nT) has been quantitatively estimated

      \item The November 2003 extreme magnetic storm is used as the benchmark event and for model performance assessment when predicting drag effects

      \item Interplay between storm-time duration and minimum Dst and Dst-like values determine the severity of satellite drag effects in low-Earth orbit 

    \end{itemize}

    \section*{Plain Language Summary}

      We investigate drag effects on satellites orbiting Earth in its upper atmosphere during magnetic storms caused by the impacts of solar superstorms. During magnetic storms, the upper atmosphere is heated and expands upwards, resulting in increased drag forces on satellites flying in those regions. Enhanced drag effects directly impact operations of such spacecraft, for instance, orbital tracking and predictions, maneuvers, and lifetime maintenance. The U.S. Federal Government has recognized space weather phenomena as natural hazards, and the understanding of their consequences, particularly during extreme circumstances, is of paramount importance. The very extreme events, here termed magnetic superstorms, occurred before the space era when no in-situ observations of the atmospheric density are available. Therefore, we use an empirical model to estimate drag from these historical events. Results generally show that the most extreme events drive the most severe effects. Additionally, we point out that another storm feature, its time duration, can play a significant role in enhancing drag. Therefore, we argue that space weather forecasters should be aware of events with long duration, particularly the ones caused by sequential impacts of solar disturbances on the Earth’s magnetic field, when predicting and forecasting the subsequent drag effects on satellites in the upper atmosphere.

    \section{Introduction}\label{introduction}

      Magnetic storms are global phenomena that occur due to the interaction of solar perturbations with the Earth's magnetosphere \citep{Gonzalez1994}. The most intense and severe magnetic storms are commonly caused by coronal mass ejections (CMEs) \citep{Gonzalez1994,Daglis1999,Balan2014}. CMEs usually have a shock at their leading edge that is promptly followed by a sheath and a magnetic cloud \citep{Gonzalez1994,Balan2014,Kilpua2019b}. Extreme magnetic storms are caused by the impact of extremely fast CMEs on the Earth's magnetosphere \citep{Tsurutani2014a}, usually associated with highly depressed values of the southward component of the interplanetary magnetic field \citep{Gonzalez1994,Daglis1999,Balan2014,Tsurutani2014a,Kilpua2019b}.\par

      Extreme space weather events like severe magnetic storms have been recognized by the U.S. Federal Government through the National Space Weather Strategy and Action Plan \citep{NSWS2015,NSWAP2015} as a natural hazard, and the need to establish benchmarks for extreme space weather events has also been recognized by the scientific community \citep[e.g.,][]{Lanzerotti2015,Jonas2017b,Riley2018}. The intensity of magnetic storms is usually measured by depletions of the ground horizontal magnetic field component recorded by magnetometers located at mid- and low-latitudes by means of the disturbance storm time (Dst) index (section \ref{dst_index}). Extremely severe events, here termed magnetic superstorms, with minimum Dst $\leq$ --500 nT, are notably rare \citep{Cliver2013,Riley2018,Vennerstrom2016,Hayakawa2019b,Chapman2020}. For instance, the March 1989 event, the only superstorm occurring during the space age \citep{Meng2019}, is well-known for the occurrence of low-latitude aurorae \citep{Allen1989,Rich1992,Pulkkinen2012} and intense geomagnetically induced currents (GICs) which caused the blackout of the Hydro-Qu\'ebec system in Canada for several hours, leading to serious economic losses \citep{Bolduc2002,Kappenman2006,Pulkkinen2017}. However, though arguably, the most extreme ground horizontal magnetic field perturbation ($\sim$ --1600 nT) on record was recorded by the Colaba station during the Carrington event of September 1859 \citep{Tsurutani2003b,Siscoe2006,Hayakawa2019b}. Since that is the only known low-latitude data set available to date, a global analysis of that storm cannot be performed \citep{Siscoe2006,Cliver2013,Hayakawa2019b,Blake2020a}. For this reason, the Carrington event is not addressed in this paper. \par

      During active times, large amounts of electromagnetic energy enter the ionosphere-thermosphere system causing the prompt thermosphere heating and upward extension due to mechanical collisions between ions and neutrals \citep[e.g.,][]{Prolss2011,Emmert2015}. This energy has access to the thermosphere primarily through high latitudes \citep{Fuller-Rowell1994,Liu2005a,Huang2014,Connor2016,Lu2016a,KalafatogluEyiguler2018}, and propagates equatorward due to the occurrence of gravity waves and wind surges \citep{Fuller-Rowell1994,Hocke1996,Bruinsma2007,Sutton2009a}. Therefore, the heating and upwelling of the thermosphere are global phenomena \citep{Richmond2000,Liu2005b,Sutton2009a}. As a result, satellites that happen to fly in those regions experience increased effects of drag forces leading to stronger orbital degradations or altitude losses \citep{Prolss2011,Prieto2014,Zesta2016b}. The understanding and control of orbital drag effects during active times can enhance predictability and forecasting of satellite tracking, reentry processes, and maintenance of satellite life times \citep{Prolss2011,Zesta2016b,Berger2020}, particularly during extreme magnetic storms \citep{Oliveira2019b}. Most of these studies have used data obtained from state-of-the-art accelerometers onboard two low-Earth orbit (LEO) satellites, namely CHAMP \citep[CHAllenge Mini‐satellite Payload;][]{Reigber2002a} and GRACE \citep[Gravity Recovery And Climate Experiment;][]{Tapley2004a}. These spacecraft were launched after 2001 (section \ref{density_section}).  \par

      The most extreme magnetic storm experienced by CHAMP and GRACE took place in November 2003 with minimum Dst = --422 nT. Consequently, there are no assessments of satellite drag in LEO during magnetic superstorms inferred from high-accuracy accelerometer data. The orbital degradations of CHAMP and GRACE associated with the November 2003 event 60 hrs through stormy times were, respectively, $\sim$ --160 m and $\sim$ --71 m \citep{Krauss2015,Oliveira2019b}, much more severe than the natural drag caused by the quiet-time backgorund density estimated by \cite{Oliveira2019b}, namely --24.11 m and --6.86 m, respectively. Hence, these are the most extreme storm-time orbital decays measured with high-quality accelerometer data. In order to empirically estimate drag effects during magnetic superstorms, standard Dst data and ground magnetometer data of historical superstorms reconstructed from historical archives are used by a thermospheric empirical model (section \ref{JB2008_section}) for density computations (section \ref{drag_framework}). These events occurred in March 1989 \citep{Allen1989,Boteler2019}, with the traditional Dst index available, September 1909 \cite{Silverman1995,Hayakawa2019a}, May 1921 \citep{Silverman2001,Hapgood2019}, and October/November 1903 \citep{Lockyer1903,Ribeiro2016}, with an alternative version to the Dst index available. These four magnetic superstorms are here examined because they are the only events with known and complete magnetograms that satisfy the threshold Dst/Dst-like $\leq$ --500 nT. The main characteristics of these storms' effects will be presented in section 3.1. Effects of storm time duration associated with minimum values of Dst and Dst-like data will be estimated and compared. As a result, this effort will improve our understanding of severe satellite orbital drag effects in LEO caused by magnetic superstorms. \par

      \section{Data, model, and a framework for orbital drag estimations}\label{data_model}

        \subsection{Disturbance storm time indices}\label{dst_index}

\begin{figure}[t]
            \centering
            \includegraphics[width=0.87\textwidth]{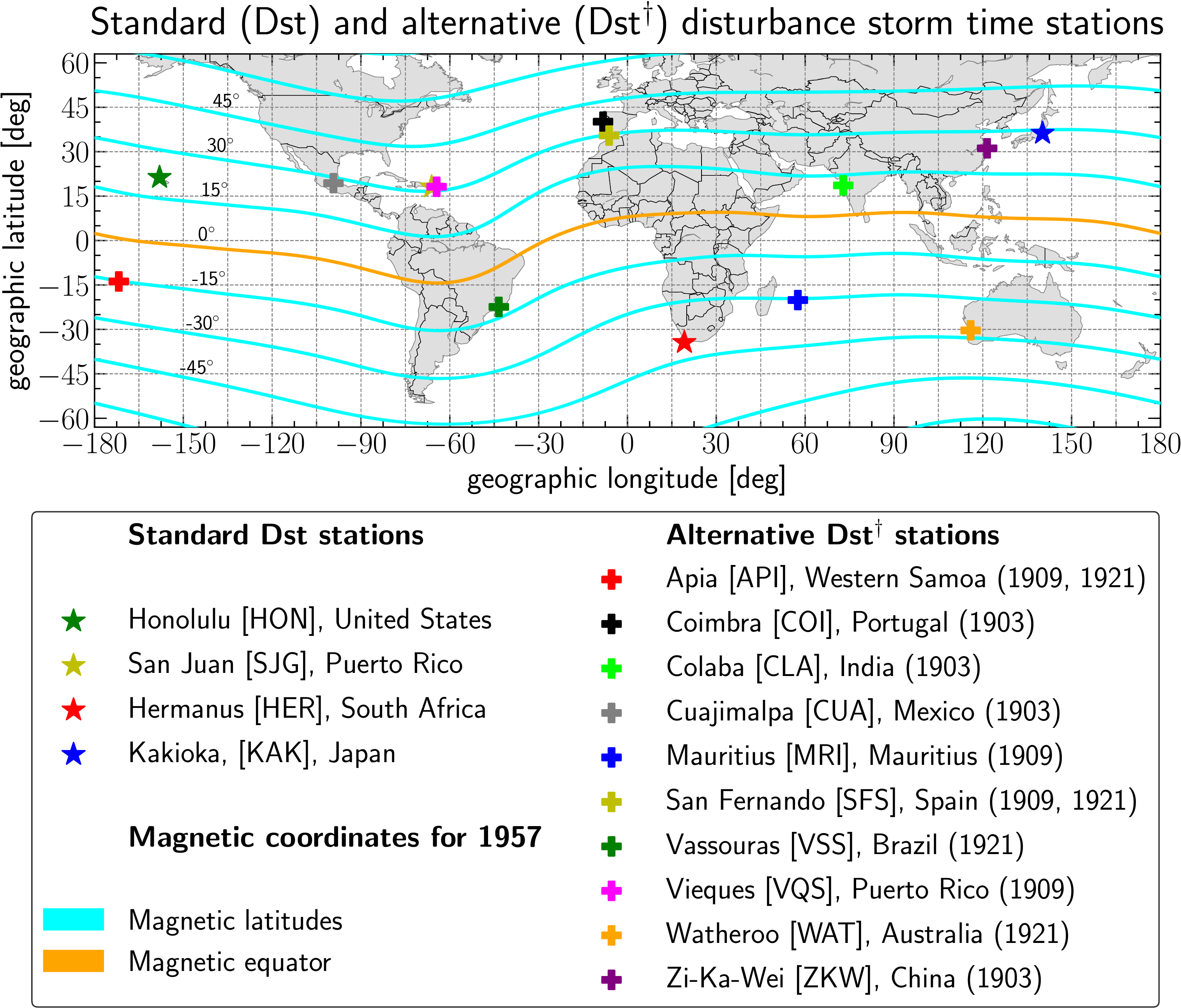}
            \caption{Geographic locations of the ground magnetometer stations that compose the standard Dst network that has been used by the \cite{WDC_Dst2015} since 1957 (colored stars), and the alternative Dst$^\dagger$ network used by \cite{Hayakawa2020a}, \cite{Love2019a}, and \cite{Love2019b} for the historical events of October/November 1903, September 1909 and May 1921 (colored crosses), respectively. Magnetic latitudes (solid cyan lines) and the magnetic equator (solid orange line) were computed by the Altitude-Adjusted Corrected Geomagnetic Coordinates Model \cite{Shepherd2014,Laundal2017} for 1957. Note that SJG is very close to VQS and therefore not clearly shown in this figure.}
            \label{stations}
          \end{figure}

          In this study, magnetic activity is represented by the Dst index provided by the \cite{WDC_Dst2015}. This 1-hr-resolution index was defined in 1957, the International Geophysical Year (IGY), as described by \cite{Sugiura1964a}. Specifically, Dst is computed by averaging latitudinally weighted horizontal magnetic field perturbations, with a background removal scheme, recorded by mid- and low-latitude stations with reasonably even longitudinal separation according to the expression

          \begin{equation}
            Dst = \frac{1}{4}\sum\limits_{i=1}^{4}\frac{\Delta H_i}{\cos\Lambda_i}\,, \quad i\mbox{ in [HON, SJG, HER, KAK]}
          \end{equation}
          where $\Delta H_i$ is the horizontal magnetic perturbation of the i-th station, and $\Lambda_i$ is the contemporary magnetic latitude of the i-th station. The colored stars in Figure \ref{stations} show the stations, with their corresponding names, abbreviations, and geographic locations, used to compute standard Dst after the IGY. \par

          \begin{equation}
            Dst = \frac{1}{4}\sum\limits_{i=1}^{4}\frac{\Delta H_i}{\cos\Lambda_i}\,, \quad i\mbox{ in [HON, SJG, HER, KAK]}
          \end{equation}
          where $\Delta H_i$ is the horizontal magnetic perturbation of the i-th station, and $\Lambda_i$ is the contemporary magnetic latitude of the i-th station. The colored stars in Figure \ref{stations} show the stations, with their corresponding names, abbreviations, and geographic locations, used to compute standard Dst after the IGY. \par

          Additionally, recent efforts have been undertaken to provide alternative (but similar) versions to the standard Dst index for historical magnetic superstorms with archival material. The events took place in October/November 1903 \citep{Hayakawa2020a}, September 1909 \citep{Love2019a}, and May 1921 \citep{Love2019b}. This alternative index, also with resolution of 1 hr, was reconstructed with data obtained from four low/mid-latitude stations, with the best possible longitudinal separation, and is represented here by Dst$^\dagger$ (Dst ``dagger"). The corresponding contemporary magnetic latitudes were computed by the authors. A background removal scheme similar to the one used to calculate Dst is used in the source papers as well. The stations used to compute Dst$^\dagger$ used in this study are shown by the colored crosses in Figure \ref{stations}. Therefore, the Dst$^\dagger$ index is given by

          \begin{equation}
            Dst^\dagger = \frac{1}{4}\sum\limits_{j=1}^{4}\frac{\Delta H_j}{\cos\Lambda_j}\,,\quad j\mbox{ in}\begin{cases}
            \text{[CLA, COI, CUA, ZKW]} &\text{for Oct/Nov 1903} \\
            \text{[API, MRI, SFS, VQS]} &\text{for Sep 1909} \\
            \text{[API, SFS, VSS, WAT]} &\text{for May 1921}
              \end{cases}
          \end{equation}

          The Dst$^\dagger$ data for the magnetic superstorms used here are available as supporting information provided by the respective references \citep{Hayakawa2020a,Love2019a,Love2019b}. Details of individual stations and magnetograms for each corresponding Dst$^\dagger$ network are provided in the source articles. \par

        \subsection{Neutral mass density data}\label{density_section}

          CHAMP and GRACE neutral mass density ($\rho$) data obtained from their respective high-accuracy accelerometers are used in this work. CHAMP was launched in 2001 at the initial altitude 456 km and orbital inclination 87.25$^\circ$. It covered each 1 hr local time in 5.5 days with orbital period 90 min. The GRACE-A and -B spacecraft were launched in 2002 at the initial altitude 500 km and orbital inclination 89.5$^\circ$. The GRACE constellation covered each 1 hr local time in 6.7 days with orbital period 95 min. GRACE-A flew $\sim$220 km ahead of GRACE-B. As discussed in \cite{Oliveira2019b}, only GRACE-A data are used, henceforth GRACE data, because GRACE-A data show higher quality than GRACE-B data. CHAMP re-entered in 2010, while GRACE re-entered in 2018. Uncertainties and calibration techniques of both missions have been discussed by many papers \citep[e.g.,][]{Bruinsma2004,Doornbos2006,Flury2008}. \par 

          The density data used in this study are normalized and intercalibrated as described in \cite{Oliveira2017c} and \cite{Zesta2019a}. Basically, the Jacchia-Bowman 2008 \citep[hereafter JB2008,][see below]{Bowman2008} empirical model computes quiet-time densities ($\rho_0$) in order to obtain the background state for the quiet thermosphere. This approach ensures that the ratio and the difference between the storm-time and quiet-time densities are as close to one ($\rho/\rho_0$ $\approx$ 1) and zero ($\rho$ -- $\rho_0$ $\approx$ 0) as possible, respectively. As a result, storm-time density enhancements can be extracted more effectively \citep{Oliveira2017c,Oliveira2019b,Zesta2019a}. \par

        \subsection{The Jacchia-Bowman 2008 (JB2008) empirical model}\label{JB2008_section}

          The first clear link between magnetic activity and satellite orbital drag effects was established by \cite{Jacchia1959}, who used Sputnik 1958$\delta1$ data to discover that its altitude significantly decayed during an extreme magnetic storm. He correctly realized that this effect occurred due to augmented density levels at the satellite's altitude. Later on, this discovery led scientists to develop thermospheric empirical models such as the Jacchia 70 model \citep[J70;][]{Jacchia1970}, the Mass Spectrometer Incoherent Scatter model series \citep[MSIS;][]{Hedin1987} which were the precursors to the Naval Research Laboratory Mass Spectrometer Incoherent Scatter Extended \citep[NRLMSISE-00;][]{Picone2002}, the Drag Temperature Model \citep[DTM2013;][]{Bruinsma2015}, the High Accuracy Satellite Drag Model \citep[HASDM,][]{Storz2005}, and more importantly for this work, the JB2008 model. A description of the JB2008 model along with other popular thermospheric empirical models has recently been provided by \cite{He2018}. \par

          The JB2008 empirical model computes thermospheric neutral mass density from a single parameter, the exospheric temperature \citep[see equation 2 in][]{Oliveira2019b}. This temperature depends on several satellite parameters such as latitude, local time, and altitude. Additionally, JB2008 uses the solar radio flux at wavelength 10.7 cm, indicated by the F10.7 index, to account for thermospheric heating due to solar UV radiation \citep{Bowman2008}. Finally, a term that depends on Dst in the exospheric temperature represents the magnetic activity contribution, but JB2008 uses the 3-hr time resolution ap index for intervals when Dst $>$ --75 nT. Dst and Dst$^\dagger$ data of the historical magnetic superstorms recorded by the stations shown in Figure \ref{stations} will be used along with LEO satellite orbital data during the event of November 2003 to estimate the subsequent drag effects. \par

        \subsection{Orbital drag computations}\label{drag_framework}

          Neutral mass densities are derived by high-accuracy accelerometers according to the drag equation \citep{Prieto2014}:
          \begin{equation}
            a_d=-\frac{1}{2}\rho C_D \frac{S}{m}V^2\,\quad V=|\vec{V}_{s/c} - \vec{V}_{wind}|\,,
          \end{equation}
          where $a_d$ is the spacecraft acceleration caused by drag forces; $\rho$ is the local thermospheric neutral mass density; $C_D$ is the drag coefficient; $S/m$ is the area-to-mass ratio; and $V$ is the relative velocity between the spacecraft velocity ($\vec{V}_{s/c}$) and the ambient neutral wind velocity $(\vec{V}_{wind})$. In this equation, all quantities are presumably known, and therefore it is solved for $\rho$ in order to yield density. However, these parameters (particularly $C_D$) can introduce significant errors in density computations \citep{Moe2005,Prieto2014,Zesta2016b}. In this study, drag coefficients computed with error mitigation methods by \cite{Sutton2009b} were used. \par

          \cite{Chen2012} provide the following expression for the computation of storm-time orbital decay rate:

          \begin{equation}\label{dadt}
            \frac{\mbox{d}a}{\mbox{d}t}=-C_D\frac{S}{m}\sqrt{GM\langle a\rangle}\Delta\rho\,,
          \end{equation}

          with $a$ being the semi-major axis of the satellite orbit here replaced by the temporal Earth's radius plus satellite altitude \citep{Oliveira2019b}, $G$ = 6.67$\times10^{-11}$ m$^3\cdot$kg$^{-1}\cdot$s$^{-2}$ the gravitational constant, $M$ = 5.972$\times10^{24}$ kg the Earth's mass, and $\Delta\rho$ the difference between the modeled storm-time and quiet-time densities. As outlined by \cite{Oliveira2019b}, the daily average of the semi-major axis $a$ is represented by $\langle a\rangle$. A comparison between the use of both $\langle a\rangle$ computation methods for a magnetically quiet day (not shown) reviewed a very minimal difference in $\mbox{d}a/\mbox{d}t$. In addition, \cite{Krauss2015} and \cite{Oliveira2019b} found the same results for the orbital decay of GRACE during the November 2003 storm. \par

          Finally, the storm-time orbital decay ($d(t)$) is computed by the sum over all $\mbox{d}a/\mbox{d}t$ values along the satellite's path for any ($t_1$, $t_2$) interval:

          \begin{equation}\label{d}
            d(t)=\int\limits_{t1}^{t2}a'(t)\mbox{d}t\,,
          \end{equation}
          where $a'(t)=\mbox{d}a(t)/\mbox{d}t$.

      \section{Results}

        \subsection{The selected magnetic superstorms}\label{superstorm_section}

\begin{figure}
            \centering
            \includegraphics[width=0.85\textwidth]{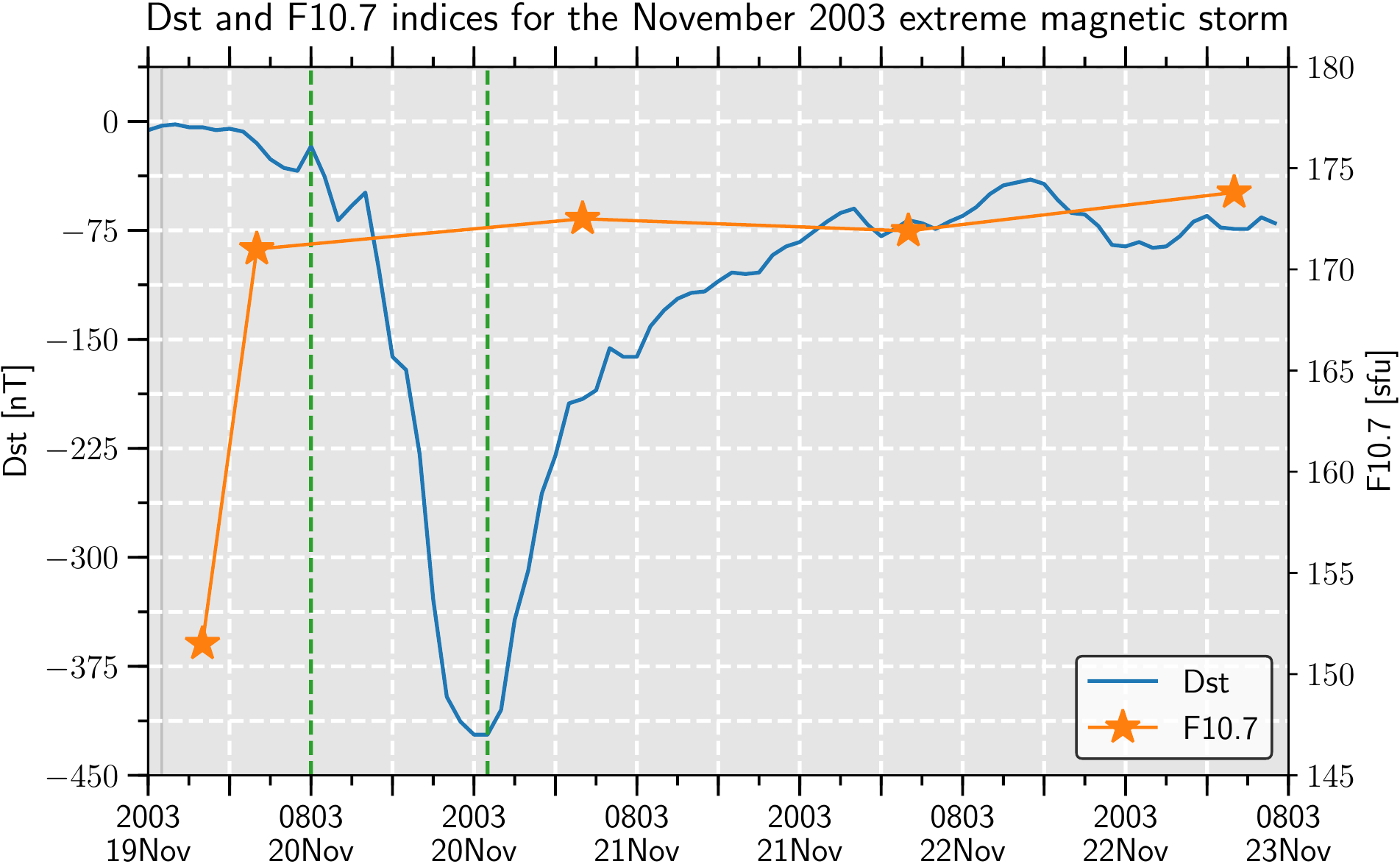}
            \caption{Dst data (blue solid line) and F10.7 data (solid orange line) for the extreme magnetic storm of November 2003. The two dashed green vertical lines indicate the 13-hr time interval between CME impact and minimum Dst value occurrence.}
            \label{f107}
          \end{figure}

          The benchmark event for the current study occurred in November 2003. That storm had minimum Dst = --422 nT, the most intense magnetic storm event with both CHAMP and GRACE neutral mass density data available. Ground magnetometer data and neutral mass density data for the GRACE satellite are shown in Figure 1 of \cite{Zesta2019a}. The solar flux F10.7 index increased from 151 sfu (solar flux units) on 19 November to 175 sfu on 23 November. The Dst and F10.7 indices for that storm are shown in Figure \ref{f107}. \par

\begin{figure}[t]
            \centering
            \includegraphics[width=0.70\textwidth]{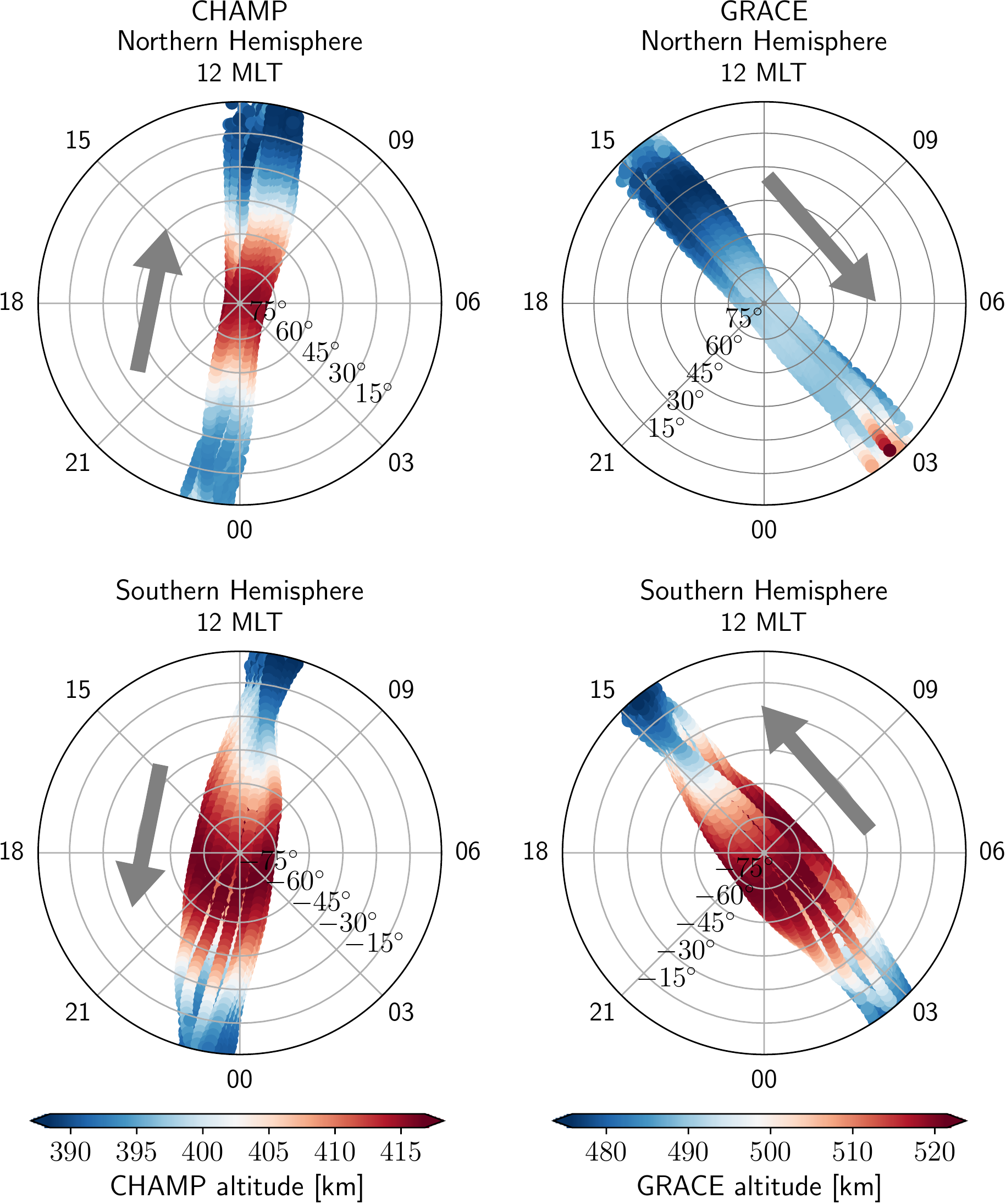}
            \caption{CHAMP (left-hand-side column) and GRACE (right-hand-side column) orbits, in magnetic coordinates (Altitude-Adjusted Corrected Geomagnetic Model coordinate system), for the northern hemisphere (top row) and southern hemisphere (bottom row). The colorbars represent the corresponding altitudes during the time interval 19-23 November 2003, the benchmark event chosen for this study. The grey arrows in all panels indicate CHAMP's and GRACE's trajectories in both hemispheres.}
            \label{sat_orbs}
          \end{figure}

          Figure \ref{sat_orbs} documents the orbits of CHAMP and GRACE in the time interval from 19 to 23 November 2003. The dial plots show orbits as a function of magnetic latitudes (MLATs) and magnetic local times (MLTs). The magnetic coordinate system used is the Altitude-Adjusted Corrected Geomagnetic Model \citep[AACGM,]{Shepherd2014,Laundal2017}. The left column shows altitudes for CHAMP, while the right column shows altitudes for GRACE. The top row indicates data for the northern hemisphere, while the bottom row indicates data for the southern hemisphere. The colorbars indicate altitudes for both satellites in the same periods. \par

          CHAMP is in a near noon-midnight orbit. The orbit altitudes of CHAMP increased at high latitudes and at the magnetic poles of both hemispheres and decreased at mid- and low-latitudes. Similar behavior is shown by GRACE whose orbits were confined within the mid-noon/dusk and mid-midnight/dawn sectors. Therefore, both spacecraft provide reasonable coverage between the day and night sectors. The altitude variations shown in Figure \ref{sat_orbs} caused by density variations at different MLATs and MLTs are mitigated by the density intercalibration method introduced by \cite{Oliveira2017c}. \par

\begin{figure}[t]
            \centering
            \includegraphics[width=1.00\textwidth]{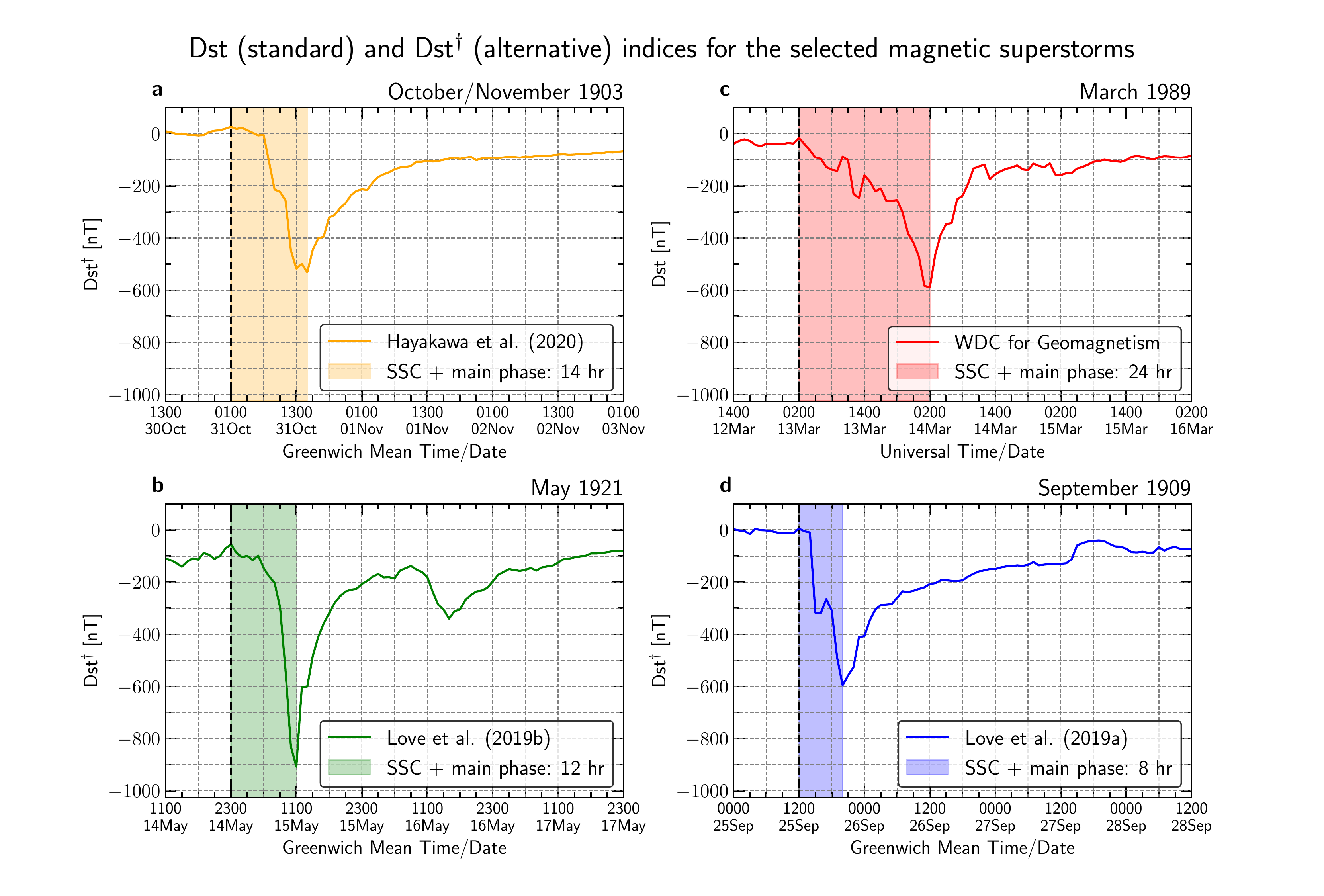}
            \caption{Ground magnetometer Dst and Dst$^\dagger$ time series, with resolution of 1 hr, for the storms of (a) October/November 1903 \citep[Dst$^\dagger$,]{Hayakawa2020a}; (b) May 1921 \citep[Dst$^\dagger$,]{Love2019b}; (c) March 1989 \citep[Dst]{WDC_Dst2015}; and (d) September 1909 \citep[Dst$^\dagger$,]{Love2019a}. The highlighted regions correspond to the time span between storm sudden commencement (SSC, vertical dashed lines) and the beginning of the storm recovery phases (minimum Dst or Dst$^\dagger$), or time duration of storm development.}
            \label{dst}
          \end{figure}

          CME leading edges are usually associated with the occurrence of positive jumps in the Dst index, while its sudden depression is associated with the arrival of CME magnetic material or sheaths \citep[e.g.,][]{Gonzalez1994,Kilpua2019b}. The first perturbation, termed storm sudden commencement (SSC), is caused by the shock compression \citep[e.g.,][]{Oliveira2018b,Shi2019b}, while the second event, termed storm main phase, is associated with strong driving of the magnetosphere via magnetic reconnection \citep[e.g.,][]{Gonzalez1994,Daglis1999,Kilpua2019b}. Examples of SSCs and storm main phases represented by the Dst and Dst$^\dagger$ indices during magnetic superstorms caused by fast CMEs are illustrated in Figure \ref{dst}. \par 
  
          Figure \ref{dst} shows ground magnetometer time series for the magnetic superstorms of (a) October/November 1903 (Dst$^\dagger$); (b) May 1921 (Dst$^\dagger$); (c) March 1989 (Dst); and (d) September 1909 (Dst$^\dagger$). Data are plotted 12 hr and 72 hr around each respective SSC (dashed vertical black lines). Times are shown as Greenwich Mean Time (GMT) for all events, except as Universal Time (UT) for the 1989 event because UTs were introduced only in 1928 \citep{Hapgood2019}. Given the similarities of UTs and GMTs, here they will be used interchangeably \citep{Hapgood2019}. The highlighted areas of each panel correspond to the time interval between SSC and minimum Dst/Dst$^\dagger$ occurrences, which also mark the beginning of the storm recovery phase. This time interval will henceforth be referred to as the storm development duration time in this paper. \par

          Panels (a) and (b) show that the 1903 event is the weakest (minimum Dst$^\dagger$ = --531 nT), whilst the 1921 event is the strongest (minimum Dst$^\dagger$ = --907 nT) amongst all events. In contrast, the development duration times of both events are almost the same, $\sim$ 14 hr and $\sim$ 12 hr, respectively. Storm strengths can be estimated by computing how fast Dst (or Dst$^\dagger$) is depressed during storm development. The average slope of Dst/Dst$^\dagger$ during the development phase is quantified by the difference of  Dst/Dst$^\dagger$ minimum minus Dst/Dst$^\dagger$ peak at SSC compression by the development time. This provides a quantifiable measure of the impactfulness of the storm, meaning that storms with very low amplitude rates are commonly associated with high geomagnetic activity \citep[e.g.,][]{Gonzalez1994}. The estimated amplitude rates are --46.4 nT/hr and --80.0 nT/hr for the October/November 1903 and May 1921 events, respectively. These numbers explain why the effects of the 1921 event, such as equatorial extent of low-latitude aurorae \citep{Chree1921,Silverman2001}, and GIC impacts on contemporary telegraph systems \citep{Kappenman2006,Hapgood2019} were more severe than the effects of the 1903 event, mostly represented by mid-latitude aurorae \citep{Page1903,Hayakawa2020a}, and local GIC impacts on contemporary telegraph systems in the United States and in the Iberian Peninsula \citep{Ribeiro2016,Hayakawa2020a}. \par    

          On the other hand, the superstorms of March 1989 and September 1909 (panels c and d) had very similar minimum values for Dst and Dst$^\dagger$, around --590 nT. However, the storm development duration of the 1989 event (24 hr) was 3 times longer than that of the 1909 event (8 hr). Consequently, the development amplitude rates of both superstorms were --23.8 nT/hr and --75.0 nT/hr, respectively. With respect to the aurorae of these events, \cite{Hayakawa2019a} estimated, based on contemporary observations, that their equatorward extent reached $\sim$ 32$^\circ$ MLAT during the 1909 superstorm, as opposed to 40$^\circ$ MLAT estimated from particle precipitation measurements by satellites during the 1989 superstorm \citep{Rich1992, Pulkkinen2012}. Intense GICs occurred during both events, with several reports of geophysical disturbances on telegraph systems in 1909 \citep{Silverman1995,Hayakawa2019a,Hapgood2019,Love2019a}, and on power transmission lines in 1989, particularly the power blackout in Qu\'ebec, Canada \citep{Allen1989,Kappenman2006,Oliveira2017d,Boteler2019}. During the 1989 event, the only event with satellite-based data amongst the four superstorms, the number of space objects ``lost" in LEO increased dramatically around periods of maximum intensity due to errors introduced by storm heating effects into tracking systems \citep{Allen1989,Joselyn1990b,Burke2018}. The left (not highlighted) part of Table \ref{properties_table} summarizes these storm properties. \par

          A comprehensive comparison of GIC effects caused by the superstorms on the contemporary ground infrastructure, i.e., telegraph systems and power grids, is a difficult task to be accomplished. However, the comparisons above show that the latitudinal extent of the auroral oval was more equatorward for the events with lower amplitude rates (May 1921 and September 1909 events). Next, the effects of these amplitude rates on storm-time orbital drag will be evaluated and compared for the four historical magnetic superstorms studied in this paper.

        \subsection{Storm-time orbital drag effects}\label{draf_effects_section}

          \subsubsection{The November 2003 extreme magnetic storm}

          Since the November 2003 magnetic storm is chosen in this work as the benchmark event, CHAMP and GRACE thermospheric neutral mass density response and the subsequent orbital drag effects for that storm are shown here, and an effort to compute errors associate with drag effects is performed. The orbital drag framework of \cite{Oliveira2019b} summarized in section \ref{drag_framework} is used for the drag computations. The Dst and F10.7 indices for the benchmark event are shown in Figure \ref{f107}. \par

          \begin{figure}
            \centering
            \includegraphics[width=0.98\textwidth]{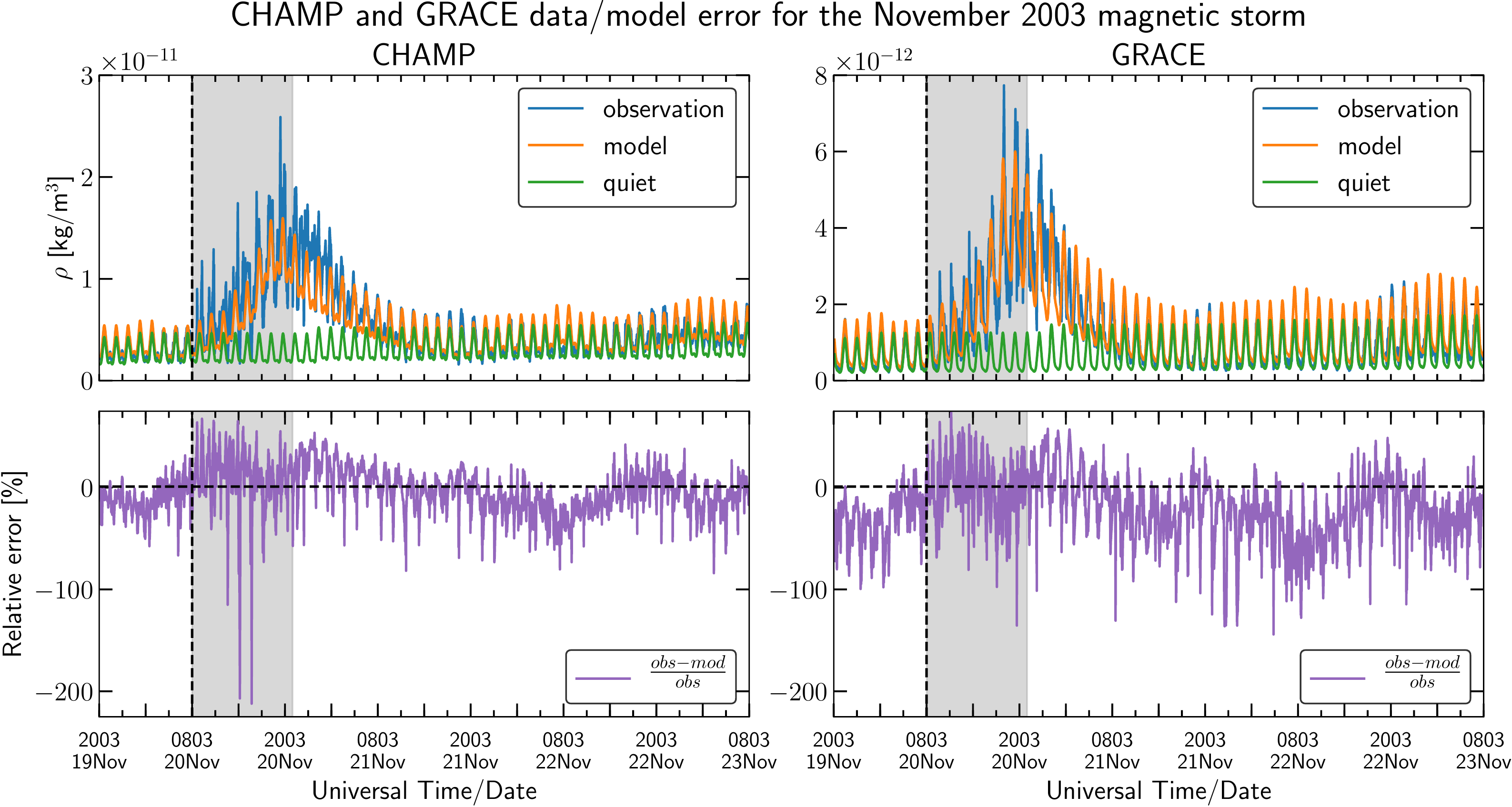}
            \caption{Top row: Observed densities by CHAMP (left) and GRACE (right) and quiet- and storm-time densities predicted by JB2008 for the November 2003 benchmark event. Bottom row: CHAMP (left) and GRACE (right) relative error between observed and modeled thermospheric densities for the same event. The grey highlighted area corresponds to the storm development time (time interval between SSC occurrence and minimum Dst value), which is 13 hrs in this case.}
            \label{error}
          \end{figure}

          Figure \ref{error} documents density observed by CHAMP (upper left) and GRACE (upper right) along with JB2008 quiet- and storm-time density predictions for the benchmark storm. The dynamics of that storm orbital effects were discussed in detail by \cite{Oliveira2019b} particularly for GRACE's case. In general, the predicted density dynamics follows CHAMP and GRACE observations quite well, but there are remarkable differences with respect to density values. Firstly, density for both satellites was highly underestimated during heating and cooling of the thermosphere, being more severe in CHAMP's case. Secondly, overestimations of JB2008 results for GRACE's orbit are higher than CHAMP's during thermospheric recovery times \citep{Oliveira2019b,Zesta2019a}. This density dynamics is reflected on the observed/predicted density relative errors shown by the solid purple lines of Figure \ref{error} in the lower left panel for CHAMP and in the lower right panel for GRACE. \par

\begin{figure*}
            \centering
            \includegraphics[width=1.00\textwidth]{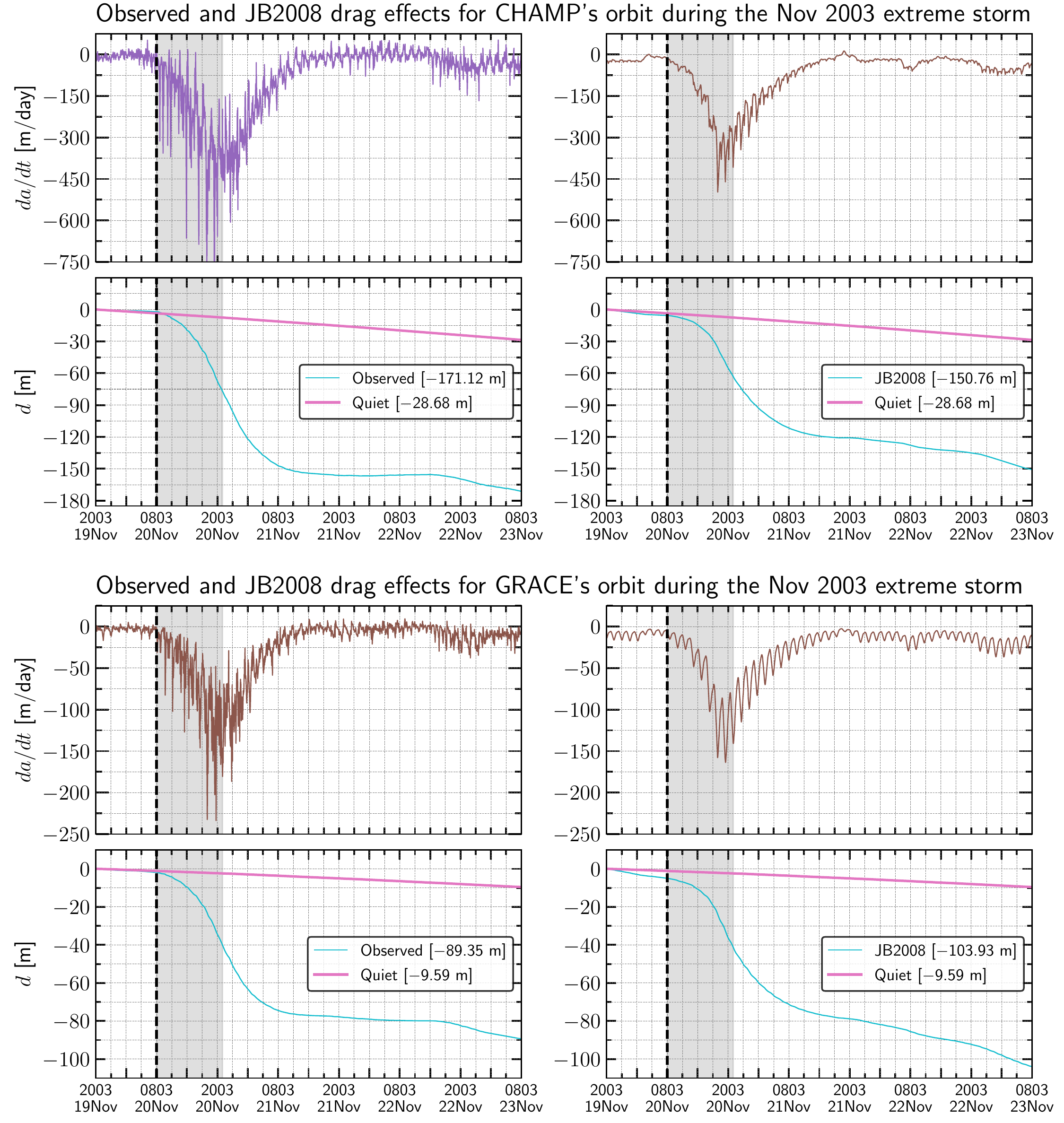}
            \caption{Top four panels: Orbital drag effects measured by CHAMP (left column) and estimated for the same orbit by JB2008 (right column) during the November 2003 extreme magnetic storm. The framework presented in section \ref{drag_framework} and introduced by \cite{Oliveira2019b} were used for the computations. The four lower panels show similar results for GRACE.}
            \label{ch_gr_drag_2003}
          \end{figure*}

          Figure \ref{ch_gr_drag_2003} shows drag results for CHAMP's and GRACE's orbits, respectively. The grey highlighted areas in all panels indicate the storm development time (13 hrs) similarly to the ones shown in Figure \ref{dst}. The odd rows of these figures show storm-time orbital decay rates ($\mbox{d}a/\mbox{d}t$, equation 4), while the even rows show storm-time orbital decay ($d$, equation 5). The left column shows observation results, while the right column shows JB2008 results. In the even rows, the magenta lines indicate the ``natural" orbital decay caused by the background density if there was no storm activity. The background density for storm-time drag computations was obtained by the method developed by \cite{Oliveira2017c}. \par

          As a result of the density dynamics shown in Figure \ref{error}, at t = 72 hrs, the storm-time orbital decays estimated for CHAMP and GRACE shown in Figure \ref{ch_gr_drag_2003} are underestimated by 13.57\% and overestimated by 16.32\%, respectively. However, the uncertainties associated with the magnetic superstorms here investigated should differ from these uncertainties for two reasons: (i) the superstorms are more intense, and (ii) the superstorms had different development times and therefore different magnetic activity during different times. These uncertainties are obtained for the most extreme magnetic storm during both CHAMP and GRACE commission times, and therefore may represent an upper limit of JB2008 uncertainties for extreme magnetic storms with high-level thermosphere neutral mass density available.

        \subsubsection{The magnetic superstorms}

\begin{figure*}
            \centering
            \includegraphics[width=1.00\textwidth]{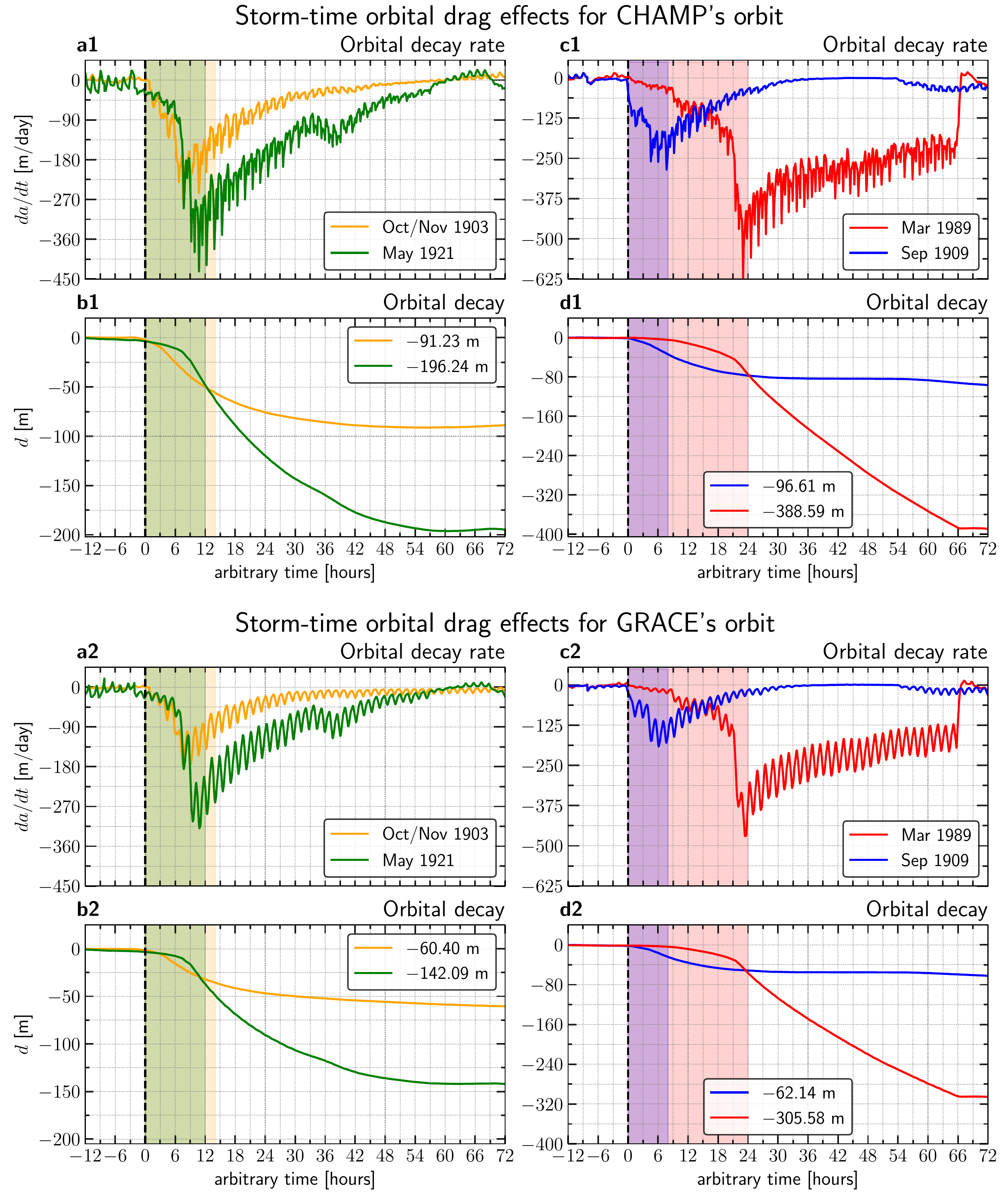}
            \caption{Satellite orbital drag predicted by JB2008 for the selected events for CHAMP's orbit (a1-d1) and GRACE's orbit (a2-d2) during the November 2003 event, but with hypothetical Dst/Dst$^\dagger$ values. Panels a1/b1 and a2/b2: $\mbox{d}a/\mbox{d}t$ and $d$ for the events in October/November 1903 (yellow lines) and May 1921 (green lines). Panels c1/d1 and c2/d2 indicate the same, but for the events in March 1989 (red lines) and September 1909 (blue lines). The highlighted areas correspond to the storm development duration, or the time interval between SSC occurrence and the end of the storm main phase (minimum Dst or Dst$^\dagger$ occurrence). These results give a sense of possible orbital decay effects during the superstorms because there are no CHAMP and/or GRACE data available during the superstorms evaluated in this paper.}
            \label{drag_effects}
          \end{figure*}

          Figure \ref{drag_effects} shows results of storm-time satellite orbital drag effects estimated according to the framework presented in section \ref{drag_framework} but for the magnetic superstorms. The computations are performed for the orbits of CHAMP and GRACE (Figure \ref{sat_orbs}), with the orbital parameters the satellites had during the November 2003 storm. The sample CHAMP- and GRACE-like satellites are flown through an upper atmosphere produced by the JB2008 model for Dst/Dst$^\dagger$ of the superstorms of Figure \ref{dst}. All solar indices are kept the same, as those of the benchmark storm. For the sake of comparisons, results are plotted as a function of arbitrary times (GMT/UT) 12 hr before and 72 hr after the SSC onset as seen in Figure \ref{dst}. The dashed vertical black lines (t = 0) indicate the times of SSC occurrence, while the highlighted areas correspond to the storm development duration as shown in Figure \ref{dst} for each corresponding storm. \par

          \begin{table}
            \centering
            \footnotesize
            \caption{Summary of the properties of the magnetic superstorms (non-highlighted area) and subsequent orbital drag results (highlighted area) shown in Figures \ref{dst} and \ref{drag_effects}, respectively. The same is shown for the benchmark event (bottom rows).}
            \begin{tabular}{c c c c c c c c}
            \hline
        \multicolumn{5}{c}{Magnetic superstorm properties} & \multicolumn{3}{c}{Orbital drag effects} \\
        \hline
        Storm & SSC & Min & Development & Amplitude & Satellite & Min & Min \\
        Month &  GMT/UT & Dst/Dst$^\dagger$ & duration$^\mathrm{b}$ & Rate$^\mathrm{c}$ & Name & $\mbox{d}a/\mbox{d}t$ & $d$ \\
         and year & (Day)$^\mathrm{a}$ & [nT] & [hr] & [nT/hr] & & [m/day] & [m] \\
        \hline
        \multirow{2}{*}{Oct/Nov 1903} & \multirow{2}{*}{0100(31)} & \multirow{2}{*}{--531} & \multirow{2}{*}{14} & \multirow{2}{*}{--46.4} & \cellcolor{gray!10}CHAMP & \cellcolor{gray!10}--272.23 & \cellcolor{gray!10}--91.23 \\
                                                & & & & & \cellcolor{gray!40}GRACE & \cellcolor{gray!40}--178.83 & \cellcolor{gray!40}--60.40 \\
        \hline
        \multirow{2}{*}{May 1921} & \multirow{2}{*}{2300(14)} & \multirow{2}{*}{--907} & \multirow{2}{*}{12} & \multirow{2}{*}{--80.0} & \cellcolor{gray!10}CHAMP & \cellcolor{gray!10}--432.98 & \cellcolor{gray!10}--196.24 \\
                                                & & & & & \cellcolor{gray!40}GRACE & \cellcolor{gray!40}--319.43 & \cellcolor{gray!40}--142.09 \\
        \hline
        \multirow{2}{*}{Mar 1989} & \multirow{2}{*}{0200(13)} & \multirow{2}{*}{--589} & \multirow{2}{*}{24} & \multirow{2}{*}{--23.8} & \cellcolor{gray!10}CHAMP & \cellcolor{gray!10}--621.29 & \cellcolor{gray!10}--388.59 \\
                                                & & & & & \cellcolor{gray!40}GRACE & \cellcolor{gray!40}--469.95 & \cellcolor{gray!40}--305.58 \\
        \hline
        \multirow{2}{*}{Sep 1909} & \multirow{2}{*}{1200(25)} & \multirow{2}{*}{--595} & \multirow{2}{*}{8} & \multirow{2}{*}{--75.0} & \cellcolor{gray!10}CHAMP & \cellcolor{gray!10}--285.14 & \cellcolor{gray!10}--96.61 \\
                                                & & & & & \cellcolor{gray!40}GRACE & \cellcolor{gray!40}--191.25 & \cellcolor{gray!40}--62.14 \\
        \hline
        \multicolumn{5}{c}{Benchmark storm properties} & \multicolumn{3}{c}{Orbital drag effects} \\
        \hline
        \multirow{2}{*}{Nov 2003} & \multirow{2}{*}{0900(20)} & \multirow{2}{*}{--422} & \multirow{2}{*}{13 } & \multirow{2}{*}{--33.8} & \cellcolor{gray!10}CHAMP$^\mathrm{d}$ & \cellcolor{gray!10}--752.43 & \cellcolor{gray!10}--171.22 \\
                                                & & & & & \cellcolor{gray!40}GRACE$^\mathrm{e}$ & \cellcolor{gray!40}--233.75 & \cellcolor{gray!40}--89.35 \\
        \hline
        \multicolumn{8}{l}{$^\mathrm{a}$ \footnotesize{Greenwich Mean Time or Universal Time and Day of Storm Sudden Commencement (SSC).}}\\
        \multicolumn{8}{l}{$^\mathrm{b}$ \footnotesize{Time between SSC and minimum Dst/Dst$^\dagger$ occurrence.}} \\
        \multicolumn{8}{l}{$^\mathrm{c}$ d(Dst/Dst$^\dagger$)/dt} \\
        \multicolumn{8}{l}{$^\mathrm{d}$ Mean altitude of CHAMP: 399.30 km} \\
        \multicolumn{8}{l}{$^\mathrm{e}$ Mean altitude of GRACE: 490.10 km}
          \end{tabular}
          \label{properties_table}
        \end{table}

          The top 4 panels of Figure \ref{drag_effects} (a1-d1) show results for CHAMP's orbit, while the bottom 4 panels (a2-d2) show results for GRACE's orbit. Panels a1 and a2 show storm-time orbital decay rates (equation \ref{dadt}) computed for the October/November 1903 superstorm (yellow line) and May 1921 superstorm (green line) for CHAMP and GRACE, respectively. Both events had approximately the same development times and very different intensities (Table \ref{properties_table}). The same is shown in panels c1 (CHAMP) and c2 (GRACE) for the superstorms of March 1989 (red line) and September 1909 (blue line). In this case, the storms had very similar intensities, but different development durations (Table \ref{properties_table}). The storm-time orbital degradation (equation \ref{d}), is shown for CHAMP (panels b1 and d1) and GRACE (panels b2 and d2). The same colors used to represent $\mbox{d}a/\mbox{d}t$ results in panels a1/c1 and a2/c2 above are used to represent $d$ results in panels b1/d1 and b2/d2. \par

          Figure \ref{drag_effects}a1 shows that $\mbox{d}a/\mbox{d}t$ values during October/November 1903 for CHAMP were very close to zero before CME impact. On the other hand, $\mbox{d}a/\mbox{d}t$ values preceding the stormy period of May 1921 show some variations (meaning $\Delta\rho$ is not necessarily close to zero), presumably linked to high magnetic activity shown by ground magnetometer data during the same pre-storm period \citep{Love2019a,Hapgood2019}. CHAMP $\mbox{d}a/\mbox{d}t$ values for the 1921 event decreased faster in comparison to minimum $\mbox{d}a/\mbox{d}t$ values for the 1903 event. Similar orbital drag dynamics is observed for GRACE (a2), but the absolute values of the drag response are smaller (Table \ref{properties_table}) because GRACE operated at higher altitudes in comparison to CHAMP \citep{Krauss2018,Oliveira2019b}. The $\mbox{d}a/\mbox{d}t$ results for CHAMP and GRACE are summarized in Table \ref{properties_table}. \par

        For the same pair of storms, the storm-time orbital degradations of CHAMP (panel b1) at the end of 72 hr after CME impact were --91.23 m and --196.24 m for both events, respectively. The same estimated results for GRACE (b2) are --60.40 m (1903) and --142.09 m (1921). Comparatively, the percentual difference between drag effects during both superstorms for CHAMP (115.10\%) are higher than the percentual difference of the superstorm intensites (70.81\%) most likely because the magnetosphere was hit by another CME on 16 May 1921 \citep[Figure \ref{dst};]{Love2019b}, leading to an additional magnetosphere energization during its recovery, which in turn impacted drag effects. Similarly, the orbital drag relative difference is higher in the case of GRACE (135.25\%), when compared with the case of CHAMP. As suggested by \cite[][Figure 10]{Oliveira2019b}, this is presumably due to the interplay between heating propagation from auroral-to-equatorial latitudes and (possibly) the direct uplift of neutrals at low and equatorial latitudes more evident at altitudes higher than 400 km \cite{Tsurutani2007}. \par 

        In summary, the main features that arise from the comparison between these events are: (i) CHAMP and GRACE decayed faster during the most intense event (1921) due to its sharper negative excursion of the Dst$^\dagger$ index and lower amplitude rate (Figure \ref{dst}a and b; Table \ref{properties_table}); and (ii) the relative differences between $d$ for both events do not closely follow the relative differences between minimum Dst$^\dagger$ values. This is likely the case because the magnetosphere was struck by another CME during its recovery, increasing the magnetospheric activity which in turn affected the subsequent orbital drag effects. Tables \ref{properties_table} and \ref{comparisons} summarize these results. \par

        The comparisons between estimated drag effects for the March 1989 and September 1909 superstorms are remarkably different. These events had very similar strengths (similar minimum Dst and Dst$^\dagger$ values), but their development times were quite distinct. Figure \ref{drag_effects}c1 shows that 1909 CHAMP $\mbox{d}a/\mbox{d}t$ values could have shown a very sharp negative excursion after CME impact, which follows very closely the same feature in the Dst$^\dagger$ index (Figure \ref{dst}d). The minimum $\mbox{d}a/\mbox{d}t$ value (--285.14 m/day) for the September 1909 superstorm was reached shortly before minimum Dst$^\dagger$. On the other hand, the March 1989 drag effects are quite different, since $\mbox{d}a/\mbox{d}t$ decreased more slowly in comparison to the former case due to the differences in storm development amplitude rates. This is explained by the fact that the magnetosphere was most likely struck by multiple CMEs while the storm main phase was developing \citep{Fuji1992,Lakhina2016,Boteler2019}. Similarly to the 1909 case, the minimum $\mbox{d}a/\mbox{d}t$ value (--621.29 m/day) occurred shortly before minimum Dst occurrence. The thermosphere recovery of the 1989 superstorm took longer than the thermosphere recovery of the 1909 superstorm, most likely because the magnetosphere was hit yet by more CMEs shortly after the beginning of the magnetosphere recovery (Figure \ref{dst}c). A similar behavior is shown by the GRACE results, panel c2, but with smaller absolute values due to higher GRACE altitudes. The relative differences between $\mbox{d}a/\mbox{d}t$ peak values of CHAMP and GRACE for both superstorms are 117.30\% and 145.73\%, even though both events had approximately the same minimum Dst and Dst$^\dagger$ values and very different storm development durations and amplitude rates. \par

        \begin{table}
      \centering
      \footnotesize
      \caption{Comparisons between magnetic superstorm intensity and satellite orbital drag severity for the magnetic superstorms predicted by JB2008 in this study.}
      \begin{tabular}{l c c c c c c}
        \hline
        Magnetic & Comparisons between &  \multicolumn{4}{c}{Relative differences of drag effects [\%]} \\
        Superstorm & Superstorm & \multicolumn{2}{c}{CHAMP} & \multicolumn{2}{c}{GRACE} \\
        Month/Year &  intensities and durations & $\mbox{d}a/\mbox{d}t$ & $d$ & $\mbox{d}a/\mbox{d}t$ & d \\
        \hline
        Oct/Nov 1903 & May 1921 is 70.81\% stronger & \multirow{2}{*}{59.05} & \multirow{2}{*}{115.10} & \multirow{2}{*}{78.62} & \multirow{2}{*}{135.25} \\
        May 1921 & Nearly the same durations & & & & \\
        \hline
        Sep 1909 & March 1989 is 3 times longer & \multirow{2}{*}{117.30} & \multirow{2}{*}{302.22} & \multirow{2}{*}{145.73} & \multirow{2}{*}{391.76} \\
        Mar 1989 & Nearly the same intensities & & & & \\
        \hline
        May 1921 & March 1989 is 2 times longer & \multirow{2}{*}{43.49$^\mathrm{a}$} & \multirow{2}{*}{98.02} & \multirow{2}{*}{47.12} & \multirow{2}{*}{115.06} \\
        Mar 1989 & May 1921 is 53.98\% stronger & & & & \\
        \hline
              \multicolumn{7}{l}{$^\mathrm{a}$ Percentual differences between more severe (March 1989) with respect to less severe} \\
              \multicolumn{7}{l}{(May 1921) drag effects} 
      \end{tabular}
      \label{comparisons}
    \end{table}

        Now the storm-time orbital degradations in both cases are evaluated. Figure \ref{drag_effects}d1 shows that CHAMP $d$ decreased faster during the main phase of the 1909 event, reaching values near its minimum value around the beginning of storm recovery. This is a typical feature of drag effects triggered by a storm caused by an isolated CME \citep{Krauss2015,Krauss2018,Oliveira2019b}. Conversely, CHAMP's orbital degradation decreased more dramatically during the recovery of the 1989 superstorm. These drag effects correlate well with a very sharp negative excursion presented by the Dst index, which is also directly related with the occurrence of low-latitude aurorae and very intense GICs around the world \citep{Allen1989,Kappenman2006,Hayakawa2019a}. This time also coincides with the loss of orbital control of several objects in LEO as shown by satellite-based data \citep{Allen1989,Joselyn1990b,Burke2018}. The storm-time orbital decays for the 1909 and 1989 events are --96.61 m and --388.59 m for CHAMP and --62.14 m and --305.58 m for GRACE. Their relative differences are 302.22\% and 391.76\%, closely following the proportion of storm time developments in the case of CHAMP. Taking into consideration that both superstorms were almost equally intense, these results show that the storm time duration can play a major role in driving orbital drag effects. Note also that relative differences are higher in the case of GRACE, most likely explained by the reasons suggested by \cite{Oliveira2019b} as mentioned before. \par

        Another striking difference concerning minimum Dst and Dst$^\dagger$ values, storm development duration and subsequent amplitude rate impacts arises from the comparison between the May 1921 and March 1989 superstorms. The 1921 event was more than 50\% stronger than the 1989 event, but active times during the latter lasted twice longer. The storm-time orbital decay for the March 1989 event was nearly twice more severe than the May 1921 event in both CHAMP's and GRACE's cases (Figure \ref{drag_effects} and Tables \ref{properties_table} and \ref{comparisons}). These results clearly reveal that a long-lasting magnetic superstorm can drive much more severe drag effects in comparison to a short-lasting, even stronger, superstorm. Tables \ref{properties_table} and \ref{comparisons} summarize the main results discussed in sections \ref{superstorm_section} and \ref{draf_effects_section}. \par

        \begin{table}
      \centering
      \caption{Storm-time orbital decay for the magnetic superstorms corrected against the November 2003 benchmark event.}
      \begin{tabular}{l l c c c c}
        \hline
        Satellite & orbital & \multicolumn{4}{c}{Superstorm month/year} \\
        name & decay & Oct/Nov 1903 & Sep 1909 & May 1921 & Mar 1989 \\
        \hline
        \multirow{2}{*}{CHAMP$^\mathrm{a}$} & $d$ [m] (model) & --91.23 & ---96.61 & --196.24 & --388.59 \\
                               & \cellcolor{gray!20}$d$ [m] (corrected) &\cellcolor{gray!20} --103.61 &\cellcolor{gray!20} --109.72 &\cellcolor{gray!20} --222.87 &\cellcolor{gray!20} --441.32 \\
        \hline
        \multirow{2}{*}{GRACE$^\mathrm{b}$} & $d$ [m] (model) & --60.40 & --62.14 & --142.09 & --305.58 \\
                               & \cellcolor{gray!20}$d$ [m] (corrected)& \cellcolor{gray!20}--50.54 & \cellcolor{gray!20}--52.00 & \cellcolor{gray!20}--118.90 & \cellcolor{gray!20}--255.71 \\
        \hline
        \multicolumn{6}{l}{$^\mathrm{a}$ \footnotesize{Underestimation of 13.57\%.}}\\
        \multicolumn{6}{l}{$^\mathrm{b}$ \footnotesize{Overestimation of 16.32\%.}} \\
      \end{tabular}
      \label{table3}
    \end{table}

        The results presented so far correspond to the storm-time orbital decay values estimated by JB2008. Furthermore, the uncertainties computed for CHAMP's and GRACE's orbital drag effects during November 2003 (section 3.2.1) can be used to obtain more realistic drag results. Results are shown in Table \ref{table3}, where white cells show model results, whereas grey cells show corrected results. In these new computations, only assumptions on overall error levels (at t = 72 hrs) were used since realistic errors cannot be obtained for the different superstorms because there are neither CHAMP nor GRACE density data available during these superstorm times. \par

        There are no solar wind nor interplanetary magnetic field data available for the magnetic superstorms discussed in this paper. Furthermore, it is important to emphasize that our statements concerning CME impacts are supported by our current knowledge of the underlying science: intense magnetic storms, particularly extreme events, are usually caused by CMEs \citep{Gonzalez1994,Daglis1999,Balan2014,Tsurutani2014a,Lakhina2016,Kilpua2019b}. \par

        \section{Discussion and conclusion}

          Extreme magnetic storms (minimum Dst $\leq$ --250 nT) are very rare. Only 39 extreme events have taken place since the beginning of the space era \citep{Meng2019}, while only 7 extreme events were observed by CHAMP and GRACE \citep{Oliveira2019b,Zesta2019a}. Additionally, only one magnetic superstorm (minimum Dst $\leq$ --500 nT) occurred since 1957, while none were ever observed by either CHAMP or GRACE. Therefore, current knowledge of thermospheric mass density response to magnetic supersotorms and the subsequent storm-time drag effects are very limited. Then, in order to estimate these effects, 4 historical magnetic superstorms with complete magnetograms were selected: one with standard Dst data (March 1989), and 3 with Dst$^\dagger$ (Dst-like) data occurring on October/November 1903 \citep{Hayakawa2020a}, September 1909 \citep{Love2019a}, and May 1921 \citep{Love2019b}. These Dst and Dst$^\dagger$ data were used as input data for the JB2008 thermosphric empirical model for density computations. The extreme magnetic storm of November 2003 (minimum Dst = --422 nT), the most extreme event during CHAMP's and GRACE's commission times, at the altitudes $\sim$400 km and $\sim$490 km, respectively, was used as the benchmark event. The orbital drag framework provided by \cite{Oliveira2019b} was used for drag estimations. \par

    First, two events with different intensities but with approximately the same storm development times were compared (October/November 1903 and May 1921). Although the 1921 superstorm was $\sim$ 70\% stronger than the 1903 superstorm, the drag effects in the former were up to 135\% more severe than the effects in the latter (GRACE's case). This is attributed to the likely impact of another CME during the recovery phase of the 1921 superstorm. Second, the other pair of superstorms, with very similar strengths, but with the September 1909 storm development being 3 times shorter than the March 1989 storm development, were compared. Results show that the relative difference of the storm-time orbital degradation for the 1989 event was about 400\% higher than the 1909 event (GRACE's case). This is explained by the likely impacts of several CMEs on the magnetosphere during the main and recovery phases of the March 1989 superstorm \citep{Fuji1992,Lakhina2016,Boteler2019}. Therefore, as opposed to latitudinal extent of aurorae, a superstorm with a smaller amplitude rate (absolute value) can cause more detrimental effects on orbital drag in comparison to an even stronger superstorm that develops faster (larger absolute value of amplitude rate). The CHAMP and GRACE storm-time orbital decays as predicted by JB2008 and corrected by errors obtained during the November 2003 event (Table \ref{table3}) are much more severe than the orbital degradation due to the background densities during the benchmark storm shown in Figure \ref{ch_gr_drag_2003} for CHAMP (--28.68 m) and GRACE (--9.59 m). For example, results for the March 1989 event show that the CHAMP storm-time orbital decay was estimated to be $\sim$--441.32 m: such value has never been measured by a LEO spacecraft with high-level accelerometers. Therefore, these results set a new basis for these effects. Despite the fact that these effects can have significant error levels particularly during the storm recovery phases due to the lack of nitric oxide cooling effects in the model \citep{Mlynczak2003,Bowman2008,Knipp2017a,Oliveira2019b,Zesta2019a}, these results reveal the comparative roles of time durations and strengths of magnetic superstorms in controlling drag effects. 

    The results of this work clearly show that multiple CME impacts on the Earth's magnetosphere (as in the March 1989 superstorm), particularly occurring during active times, can largely enhance satellite orbital drag due to long and sustained storm times. These drag effects can be more severe when compared to drag effects during storms caused by a single CME leading to even more intense storms, but lasting shorter. Therefore, orbital drag forecasters should be aware of potential impacts of several CMEs on the terrestrial magnetosphere during ongoing magnetic storms \citep[e.g.,][and many references therein]{Zhao2014b}. Additionally, different thermospheric empirical models should produce different results, with JB2008 outperforming NRLMSISE-00 and HASDM outperforming JB2008 \cite{Bowman2008}, but with DTM2013 outperforming JB2008 \citep{Bruinsma2015}. In a future work, simulation results using different models of tens of historical severe and extreme magnetic storms, with minimum Dst $\leq$ --250 nT \citep{Meng2019,Oliveira2019b,Zesta2019a,Chapman2020,Hayakawa2020b}, will be statistically studied. 

     \section*{Acknowledgments}

      D.M.O. acknowledges the financial support provided by NASA through the grant HISFM18-HIF (Heliophysics Innovation Fund). E.Z. was supported by the NASA Heliophysics Internal Scientist Funding Model through the grants HISFM18-0009, HISFM18-0006 and HISFM18-HIF. H.H. has been supported by the JSPS grant-in-aids JP15H05816 (PI: S. Yoden), JP17J06954 (PI: H. Hayakawa). HH was also partly funded by the Institute for Advanced Researches of Nagoya University and the research grant for Exploratory Research on Sustainable Humanosphere Science from Research Institute for Sustainable Humanosphere (RISH) of Kyoto University. A.B is supported by the Van Allen Radiation Belt Probes mission.  The Information System and Data Center in Postdam, Germany can provide access to CHAMP data through \url{https://isdc.gfz‐potsdam.de/champ‐isdc/access‐to‐the‐champ‐data/}, and to GRACE data through \url{https://isdc.gfz‐potsdam.de/grace‐isdc/grace‐gravity‐data‐and‐documentation/}. The JB2008 code along with solar and magnetic activity data is available at \url{http://sol.spacenvironment.net/jb2008/}.

    \setlength{\bibsep}{3.3pt}


\begin{thebibliography}{}

\bibitem [\protect \citeauthoryear {%
Allen%
, Sauer%
, Frank%
\BCBL {}\ \BBA {} Reiff%
}{%
Allen%
\ \protect \BOthers {.}}{%
{\protect \APACyear {1989}}%
}]{%
Allen1989}
\APACinsertmetastar {%
Allen1989}%
\begin{APACrefauthors}%
Allen, J.%
, Sauer, H.%
, Frank, L.%
\BCBL {}\ \BBA {} Reiff, P.%
\end{APACrefauthors}%
\unskip\
\newblock
\APACrefYearMonthDay{1989}{}{}.
\newblock
{\BBOQ}\APACrefatitle {Effects of the {M}arch 1989 solar activity} {Effects of
  the {M}arch 1989 solar activity}.{\BBCQ}
\newblock
\APACjournalVolNumPages{Eos\ Transactions\ AGU}{70}{46}{1479--1488}.
\newblock
\begin{APACrefDOI} \doi{10.1029/89EO00409} \end{APACrefDOI}
\PrintBackRefs{\CurrentBib}

\bibitem [\protect \citeauthoryear {%
Balan%
\ \protect \BOthers {.}}{%
Balan%
\ \protect \BOthers {.}}{%
{\protect \APACyear {2014}}%
}]{%
Balan2014}
\APACinsertmetastar {%
Balan2014}%
\begin{APACrefauthors}%
Balan, N.%
, Skoug, R.%
, Tulasi~Ram, S.%
, Rajesh, P\BPBI K.%
, Shiokawa, K.%
, Otsuka, Y.%
, Batista, I\BPBI S.%
, Ebihara, Y.%
\BCBL {}\ \BBA {} Nakamura, T.%
\end{APACrefauthors}%
\unskip\
\newblock
\APACrefYearMonthDay{2014}{}{}.
\newblock
{\BBOQ}\APACrefatitle {{CME} front and severe space weather} {{CME} front and
  severe space weather}.{\BBCQ}
\newblock
\APACjournalVolNumPages{Journal\ of\ Geophysical\ Research:\ Space\
  Physics}{119}{12}{10,041-10,058}.
\newblock
\begin{APACrefDOI} \doi{10.1002/2014JA020151} \end{APACrefDOI}
\PrintBackRefs{\CurrentBib}

\bibitem [\protect \citeauthoryear {%
Berger%
, Holzinger%
, Sutton%
\BCBL {}\ \BBA {} Thayer%
}{%
Berger%
\ \protect \BOthers {.}}{%
{\protect \APACyear {2020}}%
}]{%
Berger2020}
\APACinsertmetastar {%
Berger2020}%
\begin{APACrefauthors}%
Berger, T.%
, Holzinger, M.%
, Sutton, E.%
\BCBL {}\ \BBA {} Thayer, J.%
\end{APACrefauthors}%
\unskip\
\newblock
\APACrefYearMonthDay{2020}{}{}.
\newblock
{\BBOQ}\APACrefatitle {{Flying Through Uncertainty}} {{Flying Through
  Uncertainty}}.{\BBCQ}
\newblock
\APACjournalVolNumPages{Space\ Weather}{18}{1}{e2019SW002373}.
\newblock
\begin{APACrefDOI} \doi{10.1029/2019SW002373} \end{APACrefDOI}
\PrintBackRefs{\CurrentBib}

\bibitem [\protect \citeauthoryear {%
Blake%
\ \protect \BOthers {.}}{%
Blake%
\ \protect \BOthers {.}}{%
{\protect \APACyear {2020}}%
}]{%
Blake2020a}
\APACinsertmetastar {%
Blake2020a}%
\begin{APACrefauthors}%
Blake, S\BPBI P.%
, Pulkkinen, A.%
, Schuck, P\BPBI W.%
, Nevanlinna, H.%
, Reale, O.%
, Veenadhari, B.%
\BCBL {}\ \BBA {} Mukherjee, S.%
\end{APACrefauthors}%
\unskip\
\newblock
\APACrefYearMonthDay{2020}{}{}.
\newblock
{\BBOQ}\APACrefatitle {{Magnetic Field Measurements from Rome during the
  August-September 1859 Storms}} {{Magnetic Field Measurements from Rome during
  the August-September 1859 Storms}}.{\BBCQ}
\newblock
\APACjournalVolNumPages{Journal\ of\ Geophysical\ Research:\ Space\
  Physics}{125}{}{}.
\newblock
\begin{APACrefDOI} \doi{10.1029/2019JA027336} \end{APACrefDOI}
\PrintBackRefs{\CurrentBib}

\bibitem [\protect \citeauthoryear {%
Bolduc%
}{%
Bolduc%
}{%
{\protect \APACyear {2002}}%
}]{%
Bolduc2002}
\APACinsertmetastar {%
Bolduc2002}%
\begin{APACrefauthors}%
Bolduc, L.%
\end{APACrefauthors}%
\unskip\
\newblock
\APACrefYearMonthDay{2002}{}{}.
\newblock
{\BBOQ}\APACrefatitle {{GIC} observations and studies in the {Hydro-Qu\'ebec}
  power system} {{GIC} observations and studies in the {Hydro-Qu\'ebec} power
  system}.{\BBCQ}
\newblock
\APACjournalVolNumPages{Journal\ of\ Atmospheric\ and\ Solar-Terrestrial\
  Physics}{64}{16}{1793-1802}.
\newblock
\begin{APACrefDOI} \doi{10.1016/S1364-6826(02)00128-1} \end{APACrefDOI}
\PrintBackRefs{\CurrentBib}

\bibitem [\protect \citeauthoryear {%
Boteler%
}{%
Boteler%
}{%
{\protect \APACyear {2019}}%
}]{%
Boteler2019}
\APACinsertmetastar {%
Boteler2019}%
\begin{APACrefauthors}%
Boteler, D\BPBI H.%
\end{APACrefauthors}%
\unskip\
\newblock
\APACrefYearMonthDay{2019}{}{}.
\newblock
{\BBOQ}\APACrefatitle {{A Twenty‐First Century View of the March 1989
  Magnetic Storm}} {{A Twenty‐First Century View of the March 1989 Magnetic
  Storm}}.{\BBCQ}
\newblock
\APACjournalVolNumPages{Space\ Weather}{17}{10}{1427-1441}.
\newblock
\begin{APACrefDOI} \doi{10.1029/2019SW002278} \end{APACrefDOI}
\PrintBackRefs{\CurrentBib}

\bibitem [\protect \citeauthoryear {%
Bowman%
\ \protect \BOthers {.}}{%
Bowman%
\ \protect \BOthers {.}}{%
{\protect \APACyear {2008}}%
}]{%
Bowman2008}
\APACinsertmetastar {%
Bowman2008}%
\begin{APACrefauthors}%
Bowman, B\BPBI R.%
, Tobiska, W\BPBI K.%
, Marcos, F\BPBI A.%
, Huang, C\BPBI Y.%
, Lin, C\BPBI S.%
\BCBL {}\ \BBA {} Burke, W\BPBI J.%
\end{APACrefauthors}%
\unskip\
\newblock
\APACrefYearMonthDay{2008}{}{}.
\newblock
{\BBOQ}\APACrefatitle {A new empirical thermospheric density model {JB2008}
  using new solar and geomagnetic indices} {A new empirical thermospheric
  density model {JB2008} using new solar and geomagnetic indices}.{\BBCQ}
\newblock
\BIn{} \APACrefbtitle {{AIAA/AAS Astrodynamics Specialist Conference, AIAA
  2008--6438}} {{AIAA/AAS Astrodynamics Specialist Conference, AIAA
  2008--6438}}\ (\BPG~1-19).
\newblock
\APACaddressPublisher{Honolulu, HI}{}.
\PrintBackRefs{\CurrentBib}

\bibitem [\protect \citeauthoryear {%
Bruinsma%
}{%
Bruinsma%
}{%
{\protect \APACyear {2015}}%
}]{%
Bruinsma2015}
\APACinsertmetastar {%
Bruinsma2015}%
\begin{APACrefauthors}%
Bruinsma, S\BPBI L.%
\end{APACrefauthors}%
\unskip\
\newblock
\APACrefYearMonthDay{2015}{}{}.
\newblock
{\BBOQ}\APACrefatitle {{The DTM-2013 thermosphere model}} {{The DTM-2013
  thermosphere model}}.{\BBCQ}
\newblock
\APACjournalVolNumPages{Journal\ of\ Space\ Weather\ and\ Space\
  Climate}{5}{A1}{}.
\newblock
\begin{APACrefDOI} \doi{10.1051/swsc/2015001} \end{APACrefDOI}
\PrintBackRefs{\CurrentBib}

\bibitem [\protect \citeauthoryear {%
Bruinsma%
\ \BBA {} Forbes%
}{%
Bruinsma%
\ \BBA {} Forbes%
}{%
{\protect \APACyear {2007}}%
}]{%
Bruinsma2007}
\APACinsertmetastar {%
Bruinsma2007}%
\begin{APACrefauthors}%
Bruinsma, S\BPBI L.%
\BCBT {}\ \BBA {} Forbes, J\BPBI M.%
\end{APACrefauthors}%
\unskip\
\newblock
\APACrefYearMonthDay{2007}{}{}.
\newblock
{\BBOQ}\APACrefatitle {Global observations of traveling atmospheric
  disturbances {(TADs)} in the thermosphere} {Global observations of traveling
  atmospheric disturbances {(TADs)} in the thermosphere}.{\BBCQ}
\newblock
\APACjournalVolNumPages{Geophysical\ Research\ Letters}{34}{L14103}{}.
\newblock
\begin{APACrefDOI} \doi{10.1029/2007GL030243} \end{APACrefDOI}
\PrintBackRefs{\CurrentBib}

\bibitem [\protect \citeauthoryear {%
Bruinsma%
, Tamagnan%
\BCBL {}\ \BBA {} Biancale%
}{%
Bruinsma%
\ \protect \BOthers {.}}{%
{\protect \APACyear {2004}}%
}]{%
Bruinsma2004}
\APACinsertmetastar {%
Bruinsma2004}%
\begin{APACrefauthors}%
Bruinsma, S\BPBI L.%
, Tamagnan, D.%
\BCBL {}\ \BBA {} Biancale, R.%
\end{APACrefauthors}%
\unskip\
\newblock
\APACrefYearMonthDay{2004}{}{}.
\newblock
{\BBOQ}\APACrefatitle {Atmospheric densities derived from {CHAMP/STAR}
  accelerometer observations} {Atmospheric densities derived from {CHAMP/STAR}
  accelerometer observations}.{\BBCQ}
\newblock
\APACjournalVolNumPages{Planetary\ and\ Space\ Science}{62}{4}{297--312}.
\newblock
\begin{APACrefDOI} \doi{10.1016/j.pss.2003.11.004} \end{APACrefDOI}
\PrintBackRefs{\CurrentBib}

\bibitem [\protect \citeauthoryear {%
Burke%
}{%
Burke%
}{%
{\protect \APACyear {2018}}%
}]{%
Burke2018}
\APACinsertmetastar {%
Burke2018}%
\begin{APACrefauthors}%
Burke, W\BPBI J.%
\end{APACrefauthors}%
\unskip\
\newblock
\APACrefYearMonthDay{2018}{}{}.
\newblock
{\BBOQ}\APACrefatitle {{Thermospheric Dynamics during the March 1989 Magnetic
  Storm}} {{Thermospheric Dynamics during the March 1989 Magnetic
  Storm}}.{\BBCQ}
\newblock
\APACjournalVolNumPages{Sun\ and\ Geosciences}{13}{2}{163-168}.
\newblock
\begin{APACrefDOI} \doi{10.31401/SunGeo.2018.02.07} \end{APACrefDOI}
\PrintBackRefs{\CurrentBib}

\bibitem [\protect \citeauthoryear {%
Chapman%
, Horne%
\BCBL {}\ \BBA {} Watkins%
}{%
Chapman%
\ \protect \BOthers {.}}{%
{\protect \APACyear {2020}}%
}]{%
Chapman2020}
\APACinsertmetastar {%
Chapman2020}%
\begin{APACrefauthors}%
Chapman, S\BPBI C.%
, Horne, R\BPBI B.%
\BCBL {}\ \BBA {} Watkins, N\BPBI W.%
\end{APACrefauthors}%
\unskip\
\newblock
\APACrefYearMonthDay{2020}{}{}.
\newblock
{\BBOQ}\APACrefatitle {Using the aa index over the last 14 solar cycles to
  characterize extreme geomagnetic activity} {Using the aa index over the last
  14 solar cycles to characterize extreme geomagnetic activity}.{\BBCQ}
\newblock
\APACjournalVolNumPages{Geophysical\ Research\ Letters}{47}{3}{e2019GL086524}.
\newblock
\begin{APACrefDOI} \doi{10.1029/2019GL086524} \end{APACrefDOI}
\PrintBackRefs{\CurrentBib}

\bibitem [\protect \citeauthoryear {%
Chen%
, Xu%
, Wang%
, Lei%
\BCBL {}\ \BBA {} Burns%
}{%
Chen%
\ \protect \BOthers {.}}{%
{\protect \APACyear {2012}}%
}]{%
Chen2012}
\APACinsertmetastar {%
Chen2012}%
\begin{APACrefauthors}%
Chen, G\BHBI m.%
, Xu, J.%
, Wang, W.%
, Lei, J.%
\BCBL {}\ \BBA {} Burns, A\BPBI G.%
\end{APACrefauthors}%
\unskip\
\newblock
\APACrefYearMonthDay{2012}{}{}.
\newblock
{\BBOQ}\APACrefatitle {A comparison of the effects of {CIR--} and
  {CME--induced} geomagnetic activity on thermospheric densities and spacecraft
  orbits: {C}ase studies} {A comparison of the effects of {CIR--} and
  {CME--induced} geomagnetic activity on thermospheric densities and spacecraft
  orbits: {C}ase studies}.{\BBCQ}
\newblock
\APACjournalVolNumPages{Journal\ of\ Geophysical\ Research}{117}{A8}{}.
\newblock
\begin{APACrefDOI} \doi{10.1029/2012JA017782} \end{APACrefDOI}
\PrintBackRefs{\CurrentBib}

\bibitem [\protect \citeauthoryear {%
Chree%
}{%
Chree%
}{%
{\protect \APACyear {1921}}%
}]{%
Chree1921}
\APACinsertmetastar {%
Chree1921}%
\begin{APACrefauthors}%
Chree, C.%
\end{APACrefauthors}%
\unskip\
\newblock
\APACrefYearMonthDay{1921}{}{}.
\newblock
{\BBOQ}\APACrefatitle {{The Magnetic Storm of 13-17 May}} {{The Magnetic Storm
  of 13-17 May}}.{\BBCQ}
\newblock
\APACjournalVolNumPages{Nature}{107}{2690}{359}.
\newblock
\begin{APACrefDOI} \doi{10.1038/107359a0} \end{APACrefDOI}
\PrintBackRefs{\CurrentBib}

\bibitem [\protect \citeauthoryear {%
Cliver%
\ \BBA {} Dietrich%
}{%
Cliver%
\ \BBA {} Dietrich%
}{%
{\protect \APACyear {2013}}%
}]{%
Cliver2013}
\APACinsertmetastar {%
Cliver2013}%
\begin{APACrefauthors}%
Cliver, E\BPBI W.%
\BCBT {}\ \BBA {} Dietrich, W\BPBI F.%
\end{APACrefauthors}%
\unskip\
\newblock
\APACrefYearMonthDay{2013}{}{}.
\newblock
{\BBOQ}\APACrefatitle {The 1859 space weather event revisited: limits of
  extreme activity} {The 1859 space weather event revisited: limits of extreme
  activity}.{\BBCQ}
\newblock
\APACjournalVolNumPages{Journal\ of\ Space\ Weather\ and\ Space\
  Climate}{3}{A31}{}.
\newblock
\begin{APACrefDOI} \doi{10.1051/swsc/2013053} \end{APACrefDOI}
\PrintBackRefs{\CurrentBib}

\bibitem [\protect \citeauthoryear {%
Connor%
\ \protect \BOthers {.}}{%
Connor%
\ \protect \BOthers {.}}{%
{\protect \APACyear {2016}}%
}]{%
Connor2016}
\APACinsertmetastar {%
Connor2016}%
\begin{APACrefauthors}%
Connor, H\BPBI K.%
, Zesta, E.%
, Fedrizzi, M.%
, Shi, Y.%
, Raeder, J.%
, Codrescu, M\BPBI V.%
\BCBL {}\ \BBA {} {Fuller-Rowell}, T\BPBI J.%
\end{APACrefauthors}%
\unskip\
\newblock
\APACrefYearMonthDay{2016}{}{}.
\newblock
{\BBOQ}\APACrefatitle {Modeling the ionosphere-thermosphere response to a
  geomagnetic storm using physics-based magnetospheric energy input:
  {OpenGGCM-CTIM} results} {Modeling the ionosphere-thermosphere response to a
  geomagnetic storm using physics-based magnetospheric energy input:
  {OpenGGCM-CTIM} results}.{\BBCQ}
\newblock
\APACjournalVolNumPages{Journal\ of\ Space\ Weather\ and\ Space\
  Climate}{6}{A25}{1--15}.
\newblock
\begin{APACrefDOI} \doi{10.1051/swsc/2016019} \end{APACrefDOI}
\PrintBackRefs{\CurrentBib}

\bibitem [\protect \citeauthoryear {%
Daglis%
, Thorne%
, Baumjohann%
\BCBL {}\ \BBA {} Orsini%
}{%
Daglis%
\ \protect \BOthers {.}}{%
{\protect \APACyear {1999}}%
}]{%
Daglis1999}
\APACinsertmetastar {%
Daglis1999}%
\begin{APACrefauthors}%
Daglis, I\BPBI A.%
, Thorne, R\BPBI M.%
, Baumjohann, W.%
\BCBL {}\ \BBA {} Orsini, S.%
\end{APACrefauthors}%
\unskip\
\newblock
\APACrefYearMonthDay{1999}{}{}.
\newblock
{\BBOQ}\APACrefatitle {{The terrestrial ring current: Origin, formation, and
  decay}} {{The terrestrial ring current: Origin, formation, and
  decay}}.{\BBCQ}
\newblock
\APACjournalVolNumPages{Reviews\ of\ Geophysics}{37}{4}{407-438}.
\newblock
\begin{APACrefDOI} \doi{10.1029/1999RG900009} \end{APACrefDOI}
\PrintBackRefs{\CurrentBib}

\bibitem [\protect \citeauthoryear {%
Doornbos%
\ \BBA {} Klinkrad%
}{%
Doornbos%
\ \BBA {} Klinkrad%
}{%
{\protect \APACyear {2006}}%
}]{%
Doornbos2006}
\APACinsertmetastar {%
Doornbos2006}%
\begin{APACrefauthors}%
Doornbos, E.%
\BCBT {}\ \BBA {} Klinkrad, H.%
\end{APACrefauthors}%
\unskip\
\newblock
\APACrefYearMonthDay{2006}{}{}.
\newblock
{\BBOQ}\APACrefatitle {Modelling of space weather effects on satellite drag}
  {Modelling of space weather effects on satellite drag}.{\BBCQ}
\newblock
\APACjournalVolNumPages{Advances\ in\ Space\ {Research}}{37}{6}{1229--1239}.
\newblock
\begin{APACrefDOI} \doi{10.1016/j.asr.2005.04.097} \end{APACrefDOI}
\PrintBackRefs{\CurrentBib}

\bibitem [\protect \citeauthoryear {%
Emmert%
}{%
Emmert%
}{%
{\protect \APACyear {2015}}%
}]{%
Emmert2015}
\APACinsertmetastar {%
Emmert2015}%
\begin{APACrefauthors}%
Emmert, J\BPBI T.%
\end{APACrefauthors}%
\unskip\
\newblock
\APACrefYearMonthDay{2015}{}{}.
\newblock
{\BBOQ}\APACrefatitle {Thermospheric mass density: {A} review} {Thermospheric
  mass density: {A} review}.{\BBCQ}
\newblock
\APACjournalVolNumPages{Advances\ in\ Space\ {Research}}{56}{5}{773--824}.
\newblock
\begin{APACrefDOI} \doi{10.1016/j.asr.2015.05.038} \end{APACrefDOI}
\PrintBackRefs{\CurrentBib}

\bibitem [\protect \citeauthoryear {%
Flury%
, Bettadpur%
\BCBL {}\ \BBA {} Tapley%
}{%
Flury%
\ \protect \BOthers {.}}{%
{\protect \APACyear {2008}}%
}]{%
Flury2008}
\APACinsertmetastar {%
Flury2008}%
\begin{APACrefauthors}%
Flury, J.%
, Bettadpur, S.%
\BCBL {}\ \BBA {} Tapley, B\BPBI D.%
\end{APACrefauthors}%
\unskip\
\newblock
\APACrefYearMonthDay{2008}{}{}.
\newblock
{\BBOQ}\APACrefatitle {{Precise accelerometry onboard the GRACE gravity field
  satellite mission}} {{Precise accelerometry onboard the GRACE gravity field
  satellite mission}}.{\BBCQ}
\newblock
\APACjournalVolNumPages{Advances\ in\ Space\ {Research}}{42}{8}{1414-1423}.
\newblock
\begin{APACrefDOI} \doi{10.1016/j.asr.2008.05.004} \end{APACrefDOI}
\PrintBackRefs{\CurrentBib}

\bibitem [\protect \citeauthoryear {%
Fujii%
\ \protect \BOthers {.}}{%
Fujii%
\ \protect \BOthers {.}}{%
{\protect \APACyear {1992}}%
}]{%
Fuji1992}
\APACinsertmetastar {%
Fuji1992}%
\begin{APACrefauthors}%
Fujii, R.%
, Fukunishi, H.%
, Kokubun, S.%
, Sugiura, M.%
, Tohyama, F.%
, Hayakawa, H.%
, Tsuruda, K.%
\BCBL {}\ \BBA {} Okada, T.%
\end{APACrefauthors}%
\unskip\
\newblock
\APACrefYearMonthDay{1992}{}{}.
\newblock
{\BBOQ}\APACrefatitle {{Field-aligned current signatures during the March
  13–14, 1989, Great Magnetic Storm}} {{Field-aligned current signatures
  during the March 13–14, 1989, Great Magnetic Storm}}.{\BBCQ}
\newblock
\APACjournalVolNumPages{Journal\ of\ Geophysical\
  Research}{97}{A7}{10703-10715}.
\newblock
\begin{APACrefDOI} \doi{10.1029/92JA00171} \end{APACrefDOI}
\PrintBackRefs{\CurrentBib}

\bibitem [\protect \citeauthoryear {%
Fuller-Rowell%
, Codrescu%
, Moffett%
\BCBL {}\ \BBA {} Quegan%
}{%
Fuller-Rowell%
\ \protect \BOthers {.}}{%
{\protect \APACyear {1994}}%
}]{%
Fuller-Rowell1994}
\APACinsertmetastar {%
Fuller-Rowell1994}%
\begin{APACrefauthors}%
Fuller-Rowell, T\BPBI J.%
, Codrescu, M\BPBI V.%
, Moffett, R\BPBI J.%
\BCBL {}\ \BBA {} Quegan, S.%
\end{APACrefauthors}%
\unskip\
\newblock
\APACrefYearMonthDay{1994}{}{}.
\newblock
{\BBOQ}\APACrefatitle {Response of the thermosphere and ionosphere to
  geomagnetic storms} {Response of the thermosphere and ionosphere to
  geomagnetic storms}.{\BBCQ}
\newblock
\APACjournalVolNumPages{Journal\ of\ Geophysical\
  Research}{99}{A3}{3893--3914}.
\newblock
\begin{APACrefDOI} \doi{10.1029/93JA02015} \end{APACrefDOI}
\PrintBackRefs{\CurrentBib}

\bibitem [\protect \citeauthoryear {%
Gonzalez%
\ \protect \BOthers {.}}{%
Gonzalez%
\ \protect \BOthers {.}}{%
{\protect \APACyear {1994}}%
}]{%
Gonzalez1994}
\APACinsertmetastar {%
Gonzalez1994}%
\begin{APACrefauthors}%
Gonzalez, W\BPBI D.%
, Joselyn, J\BPBI A.%
, Kamide, Y.%
, Kroehl, H\BPBI W.%
, Rostoker, G.%
, Tsurutani, B\BPBI T.%
\BCBL {}\ \BBA {} Vasyli\={u}nas, V\BPBI M.%
\end{APACrefauthors}%
\unskip\
\newblock
\APACrefYearMonthDay{1994}{}{}.
\newblock
{\BBOQ}\APACrefatitle {What is a geomagnetic storm?} {What is a geomagnetic
  storm?}{\BBCQ}
\newblock
\APACjournalVolNumPages{Journal\ of\ Geophysical\
  Research}{99}{A4}{5771--5792}.
\newblock
\begin{APACrefDOI} \doi{10.1029/93JA02867} \end{APACrefDOI}
\PrintBackRefs{\CurrentBib}

\bibitem [\protect \citeauthoryear {%
Hapgood%
}{%
Hapgood%
}{%
{\protect \APACyear {2019}}%
}]{%
Hapgood2019}
\APACinsertmetastar {%
Hapgood2019}%
\begin{APACrefauthors}%
Hapgood, M.%
\end{APACrefauthors}%
\unskip\
\newblock
\APACrefYearMonthDay{2019}{}{}.
\newblock
{\BBOQ}\APACrefatitle {{The Great Storm of May 1921: An Exemplar of a Dangerous
  Space Weather Event}} {{The Great Storm of May 1921: An Exemplar of a
  Dangerous Space Weather Event}}.{\BBCQ}
\newblock
\APACjournalVolNumPages{Space\ Weather}{17}{7}{950-975}.
\newblock
\begin{APACrefDOI} \doi{10.1029/2019SW002195} \end{APACrefDOI}
\PrintBackRefs{\CurrentBib}

\bibitem [\protect \citeauthoryear {%
Hayakawa%
, Ebihara%
, Cliver%
\BCBL {}\ \protect \BOthers {.}}{%
Hayakawa%
, Ebihara%
, Cliver%
\BCBL {}\ \protect \BOthers {.}}{%
{\protect \APACyear {2019}}%
}]{%
Hayakawa2019a}
\APACinsertmetastar {%
Hayakawa2019a}%
\begin{APACrefauthors}%
Hayakawa, H.%
, Ebihara, Y.%
, Cliver, E\BPBI W.%
, Hattori, K.%
, Toriumi, S.%
, Love, J\BPBI J.%
, Umemura, N.%
, Namekata, K.%
, Sakaue, T.%
, Takahashi, T.%
\BCBL {}\ \BBA {} Shibata, K.%
\end{APACrefauthors}%
\unskip\
\newblock
\APACrefYearMonthDay{2019}{}{}.
\newblock
{\BBOQ}\APACrefatitle {{The extreme space weather event in September 1909}}
  {{The extreme space weather event in September 1909}}.{\BBCQ}
\newblock
\APACjournalVolNumPages{Monthly\ Notes\ of\ the\ Royal\ Astronomical\
  Society}{484}{3}{4083-4099}.
\newblock
\begin{APACrefDOI} \doi{10.1093/mnras/sty3196} \end{APACrefDOI}
\PrintBackRefs{\CurrentBib}

\bibitem [\protect \citeauthoryear {%
Hayakawa%
, Ebihara%
\BCBL {}\ \protect \BOthers {.}}{%
Hayakawa%
, Ebihara%
\BCBL {}\ \protect \BOthers {.}}{%
{\protect \APACyear {2020}}%
}]{%
Hayakawa2020b}
\APACinsertmetastar {%
Hayakawa2020b}%
\begin{APACrefauthors}%
Hayakawa, H.%
, Ebihara, Y.%
, Pevtsov, A.%
, Bhaskar, A.%
, Karachik, N.%
\BCBL {}\ \BBA {} Oliveira, D\BPBI M.%
\end{APACrefauthors}%
\unskip\
\newblock
\APACrefYearMonthDay{2020}{}{}.
\newblock
{\BBOQ}\APACrefatitle {{Intensity and Time Series of Extreme Solar-Terrestrial
  Storm in March 1946}} {{Intensity and Time Series of Extreme
  Solar-Terrestrial Storm in March 1946}}.{\BBCQ}
\newblock
\APACjournalVolNumPages{Monthly\ Notes\ of\ the\ Royal\ Astronomical\
  Society}{}{}{}.
\newblock
\begin{APACrefDOI} \doi{10.1093/mnras/staa1508} \end{APACrefDOI}
\PrintBackRefs{\CurrentBib}

\bibitem [\protect \citeauthoryear {%
Hayakawa%
, Ebihara%
, Willis%
\BCBL {}\ \protect \BOthers {.}}{%
Hayakawa%
, Ebihara%
, Willis%
\BCBL {}\ \protect \BOthers {.}}{%
{\protect \APACyear {2019}}%
}]{%
Hayakawa2019b}
\APACinsertmetastar {%
Hayakawa2019b}%
\begin{APACrefauthors}%
Hayakawa, H.%
, Ebihara, Y.%
, Willis, D\BPBI M.%
, Toriumi, S.%
, Iju, T.%
, Hattori, K.%
, Wild, M\BPBI N.%
, Oliveira, D\BPBI M.%
, Ermolli, I.%
, Ribeiro, J\BPBI R.%
, Correia, A\BPBI P.%
, Ribeiro, A\BPBI I.%
\BCBL {}\ \BBA {} Knipp, D\BPBI J.%
\end{APACrefauthors}%
\unskip\
\newblock
\APACrefYearMonthDay{2019}{}{}.
\newblock
{\BBOQ}\APACrefatitle {{Temporal and Spatial Evolutions of a Large Sunspot
  Group and Great Auroral Storms around the Carrington Event in 1859}}
  {{Temporal and Spatial Evolutions of a Large Sunspot Group and Great Auroral
  Storms around the Carrington Event in 1859}}.{\BBCQ}
\newblock
\APACjournalVolNumPages{Space\ Weather}{17}{11}{1553-1569}.
\newblock
\begin{APACrefDOI} \doi{10.1029/2019SW002269} \end{APACrefDOI}
\PrintBackRefs{\CurrentBib}

\bibitem [\protect \citeauthoryear {%
Hayakawa%
, Ribeiro%
\BCBL {}\ \protect \BOthers {.}}{%
Hayakawa%
, Ribeiro%
\BCBL {}\ \protect \BOthers {.}}{%
{\protect \APACyear {2020}}%
}]{%
Hayakawa2020a}
\APACinsertmetastar {%
Hayakawa2020a}%
\begin{APACrefauthors}%
Hayakawa, H.%
, Ribeiro, P.%
, Vaquero, J\BPBI M.%
, Gallego, M\BPBI C.%
, Knipp, D\BPBI J.%
, Mekhaldi, F.%
, Bhaskar, A.%
, Oliveira, D\BPBI M.%
, Notsu, Y.%
, Carrasco, V\BPBI M\BPBI S.%
, Caccavari, A.%
, Veenadhari, B.%
, Mukherjee, S.%
\BCBL {}\ \BBA {} Ebihara, Y.%
\end{APACrefauthors}%
\unskip\
\newblock
\APACrefYearMonthDay{2020}{}{}.
\newblock
{\BBOQ}\APACrefatitle {{The Extreme Space Weather Event in 1903
  October/November: An Outburst from the Quiet Sun}} {{The Extreme Space
  Weather Event in 1903 October/November: An Outburst from the Quiet
  Sun}}.{\BBCQ}
\newblock
\APACjournalVolNumPages{The\ Astrophysical\ Journal\ Letters}{897}{1}{L10}.
\newblock
\begin{APACrefDOI} \doi{10.3847/2041-8213/ab6a18} \end{APACrefDOI}
\PrintBackRefs{\CurrentBib}

\bibitem [\protect \citeauthoryear {%
He%
\ \protect \BOthers {.}}{%
He%
\ \protect \BOthers {.}}{%
{\protect \APACyear {2018}}%
}]{%
He2018}
\APACinsertmetastar {%
He2018}%
\begin{APACrefauthors}%
He, C.%
, Yang, Y.%
, Carter, B.%
, Kerr, E.%
, Wu, S.%
, Deleflie, F.%
, Cai, H.%
, Zhang, K.%
, Sagni\`eres, L.%
\BCBL {}\ \BBA {} Norman, R.%
\end{APACrefauthors}%
\unskip\
\newblock
\APACrefYearMonthDay{2018}{}{}.
\newblock
{\BBOQ}\APACrefatitle {Review and comparison of empirical thermospheric mass
  density models} {Review and comparison of empirical thermospheric mass
  density models}.{\BBCQ}
\newblock
\APACjournalVolNumPages{Progress\ in\ Aerospace\ Sciences}{103}{}{31-51}.
\newblock
\begin{APACrefDOI} \doi{10.1016/j.paerosci.2018.10.003} \end{APACrefDOI}
\PrintBackRefs{\CurrentBib}

\bibitem [\protect \citeauthoryear {%
Hedin%
}{%
Hedin%
}{%
{\protect \APACyear {1987}}%
}]{%
Hedin1987}
\APACinsertmetastar {%
Hedin1987}%
\begin{APACrefauthors}%
Hedin, A\BPBI E.%
\end{APACrefauthors}%
\unskip\
\newblock
\APACrefYearMonthDay{1987}{}{}.
\newblock
{\BBOQ}\APACrefatitle {{MSIS-86 Thermospheric Model}} {{MSIS-86 Thermospheric
  Model}}.{\BBCQ}
\newblock
\APACjournalVolNumPages{Journal\ of\ Geophysical\ Research}{92}{A5}{4649-4662}.
\newblock
\begin{APACrefDOI} \doi{10.1029/JA092iA05p04649} \end{APACrefDOI}
\PrintBackRefs{\CurrentBib}

\bibitem [\protect \citeauthoryear {%
Hocke%
\ \BBA {} Schlegel%
}{%
Hocke%
\ \BBA {} Schlegel%
}{%
{\protect \APACyear {1996}}%
}]{%
Hocke1996}
\APACinsertmetastar {%
Hocke1996}%
\begin{APACrefauthors}%
Hocke, K.%
\BCBT {}\ \BBA {} Schlegel, K.%
\end{APACrefauthors}%
\unskip\
\newblock
\APACrefYearMonthDay{1996}{}{}.
\newblock
{\BBOQ}\APACrefatitle {A review of atmospheric gravity waves and travelling
  ionospheric disturbances: 1982--1995} {A review of atmospheric gravity waves
  and travelling ionospheric disturbances: 1982--1995}.{\BBCQ}
\newblock
\APACjournalVolNumPages{Annales\ Geophysicae}{14}{9}{917--940}.
\newblock
\begin{APACrefDOI} \doi{10.1007/s00585-996-0917-6} \end{APACrefDOI}
\PrintBackRefs{\CurrentBib}

\bibitem [\protect \citeauthoryear {%
Huang%
, Su%
, Sutton%
, Weimer%
\BCBL {}\ \BBA {} Davidson%
}{%
Huang%
\ \protect \BOthers {.}}{%
{\protect \APACyear {2014}}%
}]{%
Huang2014}
\APACinsertmetastar {%
Huang2014}%
\begin{APACrefauthors}%
Huang, C\BPBI Y.%
, Su, Y\BHBI J.%
, Sutton, E\BPBI K.%
, Weimer, D\BPBI R.%
\BCBL {}\ \BBA {} Davidson, R\BPBI L.%
\end{APACrefauthors}%
\unskip\
\newblock
\APACrefYearMonthDay{2014}{}{}.
\newblock
{\BBOQ}\APACrefatitle {Energy coupling during the {A}ugust 2011 magnetic storm}
  {Energy coupling during the {A}ugust 2011 magnetic storm}.{\BBCQ}
\newblock
\APACjournalVolNumPages{Journal\ of\ Geophysical\ Research:\ Space\
  Physics}{119}{2}{1219--1232}.
\newblock
\begin{APACrefDOI} \doi{10.1002/2013JA019297} \end{APACrefDOI}
\PrintBackRefs{\CurrentBib}

\bibitem [\protect \citeauthoryear {%
Jacchia%
}{%
Jacchia%
}{%
{\protect \APACyear {1959}}%
}]{%
Jacchia1959}
\APACinsertmetastar {%
Jacchia1959}%
\begin{APACrefauthors}%
Jacchia, L\BPBI G.%
\end{APACrefauthors}%
\unskip\
\newblock
\APACrefYearMonthDay{1959}{}{}.
\newblock
{\BBOQ}\APACrefatitle {Corpuscular Radiation and the Acceleration of Artificial
  Satellites} {Corpuscular radiation and the acceleration of artificial
  satellites}.{\BBCQ}
\newblock
\APACjournalVolNumPages{Nature}{183}{526}{1662--1663}.
\newblock
\begin{APACrefDOI} \doi{10.1038/1831662a0} \end{APACrefDOI}
\PrintBackRefs{\CurrentBib}

\bibitem [\protect \citeauthoryear {%
Jacchia%
}{%
Jacchia%
}{%
{\protect \APACyear {1970}}%
}]{%
Jacchia1970}
\APACinsertmetastar {%
Jacchia1970}%
\begin{APACrefauthors}%
Jacchia, L\BPBI G.%
\end{APACrefauthors}%
\unskip\
\newblock
\APACrefYearMonthDay{1970}{}{}.
\newblock
{\BBOQ}\APACrefatitle {New static models of the thermosphere and exosphere with
  empirical temperature profiles} {New static models of the thermosphere and
  exosphere with empirical temperature profiles}.{\BBCQ}
\newblock
\BIn{} \APACrefbtitle {Spec. Rep. 313.} {Spec. rep. 313.}
\newblock
\APACaddressPublisher{Cambridge, Massachusetts}{Smithson, Astrophys. Obs.}
\PrintBackRefs{\CurrentBib}

\bibitem [\protect \citeauthoryear {%
Jonas%
, Murtagh%
\BCBL {}\ \BBA {} Bonadonna%
}{%
Jonas%
\ \protect \BOthers {.}}{%
{\protect \APACyear {2017}}%
}]{%
Jonas2017b}
\APACinsertmetastar {%
Jonas2017b}%
\begin{APACrefauthors}%
Jonas, S.%
, Murtagh, W.%
\BCBL {}\ \BBA {} Bonadonna, M.%
\end{APACrefauthors}%
\unskip\
\newblock
\APACrefYearMonthDay{2017}{}{}.
\newblock
{\BBOQ}\APACrefatitle {{Released for Public Comment: Space Weather Benchmarks
  and Operations-to-Research Plan}} {{Released for Public Comment: Space
  Weather Benchmarks and Operations-to-Research Plan}}.{\BBCQ}
\newblock
\APACjournalVolNumPages{Space\ Weather}{15}{2}{282-282}.
\newblock
\begin{APACrefDOI} \doi{10.1002/2017SW001603} \end{APACrefDOI}
\PrintBackRefs{\CurrentBib}

\bibitem [\protect \citeauthoryear {%
Joselyn%
}{%
Joselyn%
}{%
{\protect \APACyear {1990}}%
}]{%
Joselyn1990b}
\APACinsertmetastar {%
Joselyn1990b}%
\begin{APACrefauthors}%
Joselyn, J\BPBI A.%
\end{APACrefauthors}%
\unskip\
\newblock
\APACrefYearMonthDay{1990}{}{}.
\newblock
{\BBOQ}\APACrefatitle {{Case study of the great geomagnetic storm of 13 March
  1989, Astrodynamics 1989}} {{Case study of the great geomagnetic storm of 13
  March 1989, Astrodynamics 1989}}.{\BBCQ}
\newblock
\BIn{} \APACrefbtitle {{Proceedings of the AAS/AIAA Astrodynamics Conference,
  Stowe, VT, August 7-10, 1989}} {{Proceedings of the AAS/AIAA Astrodynamics
  Conference, Stowe, VT, August 7-10, 1989}}\ (\BPG~745-762).
\newblock
\APACaddressPublisher{San Diego, CA}{Univelt, Inc.}
\newblock
\APACrefnote{Part 2 (A90‐46754 21‐12)}
\PrintBackRefs{\CurrentBib}

\bibitem [\protect \citeauthoryear {%
{Kalafatoglu Eyiguler}%
, Kaymaz%
, Frissell%
, Ruohoniemi%
\BCBL {}\ \BBA {} Rast\"atter%
}{%
{Kalafatoglu Eyiguler}%
\ \protect \BOthers {.}}{%
{\protect \APACyear {2018}}%
}]{%
KalafatogluEyiguler2018}
\APACinsertmetastar {%
KalafatogluEyiguler2018}%
\begin{APACrefauthors}%
{Kalafatoglu Eyiguler}, E\BPBI C.%
, Kaymaz, Z.%
, Frissell, N\BPBI A.%
, Ruohoniemi, J\BPBI M.%
\BCBL {}\ \BBA {} Rast\"atter, L.%
\end{APACrefauthors}%
\unskip\
\newblock
\APACrefYearMonthDay{2018}{}{}.
\newblock
{\BBOQ}\APACrefatitle {Investigating Upper Atmospheric {J}oule Heating Using
  Cross-Combination of Data for Two Moderate Substorm Cases} {Investigating
  upper atmospheric {J}oule heating using cross-combination of data for two
  moderate substorm cases}.{\BBCQ}
\newblock
\APACjournalVolNumPages{Space\ Weather}{16}{8}{987-1012}.
\newblock
\begin{APACrefDOI} \doi{10.1029/2018SW001956} \end{APACrefDOI}
\PrintBackRefs{\CurrentBib}

\bibitem [\protect \citeauthoryear {%
Kappenman%
}{%
Kappenman%
}{%
{\protect \APACyear {2006}}%
}]{%
Kappenman2006}
\APACinsertmetastar {%
Kappenman2006}%
\begin{APACrefauthors}%
Kappenman, J\BPBI G.%
\end{APACrefauthors}%
\unskip\
\newblock
\APACrefYearMonthDay{2006}{}{}.
\newblock
{\BBOQ}\APACrefatitle {{Great geomagnetic storms and extreme impulsive
  geomagnetic field disturbance events - An analysis of observational evidence
  including the great storm of May 1921}} {{Great geomagnetic storms and
  extreme impulsive geomagnetic field disturbance events - An analysis of
  observational evidence including the great storm of May 1921}}.{\BBCQ}
\newblock
\APACjournalVolNumPages{Advances\ in\ Space\ {Research}}{38}{2}{188-199}.
\newblock
\begin{APACrefDOI} \doi{10.1016/j.asr.2005.08.055} \end{APACrefDOI}
\PrintBackRefs{\CurrentBib}

\bibitem [\protect \citeauthoryear {%
Kilpua%
\ \protect \BOthers {.}}{%
Kilpua%
\ \protect \BOthers {.}}{%
{\protect \APACyear {2019}}%
}]{%
Kilpua2019b}
\APACinsertmetastar {%
Kilpua2019b}%
\begin{APACrefauthors}%
Kilpua, E\BPBI K\BPBI J.%
, Fontaine, D.%
, Moissard, C.%
, Ala-Lahti, M.%
, Palmerio, E.%
, Yordanova, E.%
, Good, S\BPBI W.%
, Kalliokoski, M\BPBI M\BPBI H.%
, Lumme, E.%
, Osmane, A.%
, Palmroth, M.%
\BCBL {}\ \BBA {} Turc, L.%
\end{APACrefauthors}%
\unskip\
\newblock
\APACrefYearMonthDay{2019}{}{}.
\newblock
{\BBOQ}\APACrefatitle {{Solar Wind Properties and Geospace Impact of Coronal
  Mass Ejection-Driven Sheath Regions: Variation and Driver Dependence}}
  {{Solar Wind Properties and Geospace Impact of Coronal Mass Ejection-Driven
  Sheath Regions: Variation and Driver Dependence}}.{\BBCQ}
\newblock
\APACjournalVolNumPages{Space\ Weather}{17}{8}{1257-1280}.
\newblock
\begin{APACrefDOI} \doi{10.1029/2019SW002217} \end{APACrefDOI}
\PrintBackRefs{\CurrentBib}

\bibitem [\protect \citeauthoryear {%
Knipp%
\ \protect \BOthers {.}}{%
Knipp%
\ \protect \BOthers {.}}{%
{\protect \APACyear {2017}}%
}]{%
Knipp2017a}
\APACinsertmetastar {%
Knipp2017a}%
\begin{APACrefauthors}%
Knipp, D\BPBI J.%
, Pette, D\BPBI V.%
, Kilcommons, L\BPBI M.%
, Isaacs, T\BPBI L.%
, Cruz, A\BPBI A.%
, Mlynczak, M\BPBI G.%
, Hunt, L\BPBI A.%
\BCBL {}\ \BBA {} Lin, C\BPBI Y.%
\end{APACrefauthors}%
\unskip\
\newblock
\APACrefYearMonthDay{2017}{}{}.
\newblock
{\BBOQ}\APACrefatitle {Thermospheric Nitric Oxide Response to Shock-led Storms}
  {Thermospheric nitric oxide response to shock-led storms}.{\BBCQ}
\newblock
\APACjournalVolNumPages{Space\ Weather}{15}{2}{325-342}.
\newblock
\begin{APACrefDOI} \doi{10.1002/2016SW001567} \end{APACrefDOI}
\PrintBackRefs{\CurrentBib}

\bibitem [\protect \citeauthoryear {%
Krauss%
, Temmer%
\BCBL {}\ \BBA {} Vennerstrom%
}{%
Krauss%
\ \protect \BOthers {.}}{%
{\protect \APACyear {2018}}%
}]{%
Krauss2018}
\APACinsertmetastar {%
Krauss2018}%
\begin{APACrefauthors}%
Krauss, S.%
, Temmer, M.%
\BCBL {}\ \BBA {} Vennerstrom, S.%
\end{APACrefauthors}%
\unskip\
\newblock
\APACrefYearMonthDay{2018}{}{}.
\newblock
{\BBOQ}\APACrefatitle {{Multiple satellite analysis of the Earth's thermosphere
  and interplanetary magnetic field variations due to ICME/CIR events during
  2003-2015}} {{Multiple satellite analysis of the Earth's thermosphere and
  interplanetary magnetic field variations due to ICME/CIR events during
  2003-2015}}.{\BBCQ}
\newblock
\APACjournalVolNumPages{Journal\ of\ Geophysical\ Research:\ Space\
  Physics}{123}{10}{8884-8894}.
\newblock
\begin{APACrefDOI} \doi{10.1029/2018JA025778} \end{APACrefDOI}
\PrintBackRefs{\CurrentBib}

\bibitem [\protect \citeauthoryear {%
Krauss%
, Temmer%
, Veronig%
, Baur%
\BCBL {}\ \BBA {} Lammer%
}{%
Krauss%
\ \protect \BOthers {.}}{%
{\protect \APACyear {2015}}%
}]{%
Krauss2015}
\APACinsertmetastar {%
Krauss2015}%
\begin{APACrefauthors}%
Krauss, S.%
, Temmer, M.%
, Veronig, A.%
, Baur, O.%
\BCBL {}\ \BBA {} Lammer, H.%
\end{APACrefauthors}%
\unskip\
\newblock
\APACrefYearMonthDay{2015}{}{}.
\newblock
{\BBOQ}\APACrefatitle {Thermosphere and geomagnetic response to interplanetary
  coronal mass ejections observed by {ACE} and {GRACE}: {S}tatistical results.}
  {Thermosphere and geomagnetic response to interplanetary coronal mass
  ejections observed by {ACE} and {GRACE}: {S}tatistical results.}{\BBCQ}
\newblock
\APACjournalVolNumPages{Journal\ of\ Geophysical\ Research:\ Space\
  Physics}{120}{10}{8848--8860}.
\newblock
\begin{APACrefDOI} \doi{10.1002/2015JA021702} \end{APACrefDOI}
\PrintBackRefs{\CurrentBib}

\bibitem [\protect \citeauthoryear {%
Lakhina%
\ \BBA {} Tsurutani%
}{%
Lakhina%
\ \BBA {} Tsurutani%
}{%
{\protect \APACyear {2016}}%
}]{%
Lakhina2016}
\APACinsertmetastar {%
Lakhina2016}%
\begin{APACrefauthors}%
Lakhina, G\BPBI S.%
\BCBT {}\ \BBA {} Tsurutani, B\BPBI T.%
\end{APACrefauthors}%
\unskip\
\newblock
\APACrefYearMonthDay{2016}{}{}.
\newblock
{\BBOQ}\APACrefatitle {Geomagnetic storms: {H}istorical perspective to modern
  view} {Geomagnetic storms: {H}istorical perspective to modern view}.{\BBCQ}
\newblock
\APACjournalVolNumPages{Geoscience\ Letters}{3}{5}{1--11}.
\newblock
\begin{APACrefDOI} \doi{10.1186/s40562-016-0037-4} \end{APACrefDOI}
\PrintBackRefs{\CurrentBib}

\bibitem [\protect \citeauthoryear {%
Lanzerotti%
}{%
Lanzerotti%
}{%
{\protect \APACyear {2015}}%
}]{%
Lanzerotti2015}
\APACinsertmetastar {%
Lanzerotti2015}%
\begin{APACrefauthors}%
Lanzerotti, L\BPBI J.%
\end{APACrefauthors}%
\unskip\
\newblock
\APACrefYearMonthDay{2015}{}{}.
\newblock
{\BBOQ}\APACrefatitle {{Space Weather Strategy and Action Plan: The National
  Program Is Rolled Out}} {{Space Weather Strategy and Action Plan: The
  National Program Is Rolled Out}}.{\BBCQ}
\newblock
\APACjournalVolNumPages{Space\ Weather}{13}{12}{824-825}.
\newblock
\begin{APACrefDOI}\doi{10.1002/2015SW001334}\end{APACrefDOI}
\PrintBackRefs{\CurrentBib}

\bibitem [\protect \citeauthoryear {%
Laundal%
\ \BBA {} Richmond%
}{%
Laundal%
\ \BBA {} Richmond%
}{%
{\protect \APACyear {2017}}%
}]{%
Laundal2017}
\APACinsertmetastar {%
Laundal2017}%
\begin{APACrefauthors}%
Laundal, K\BPBI M.%
\BCBT {}\ \BBA {} Richmond, A\BPBI D.%
\end{APACrefauthors}%
\unskip\
\newblock
\APACrefYearMonthDay{2017}{}{}.
\newblock
{\BBOQ}\APACrefatitle {Magnetic coordinate systems} {Magnetic coordinate
  systems}.{\BBCQ}
\newblock
\APACjournalVolNumPages{Space\ Science\ Reviews}{206}{1-4}{1--33}.
\newblock
\begin{APACrefDOI} \doi{10.1007/s11214-016-0275-y} \end{APACrefDOI}
\PrintBackRefs{\CurrentBib}

\bibitem [\protect \citeauthoryear {%
Liu%
\ \BBA {} L\"uhr%
}{%
Liu%
\ \BBA {} L\"uhr%
}{%
{\protect \APACyear {2005}}%
}]{%
Liu2005a}
\APACinsertmetastar {%
Liu2005a}%
\begin{APACrefauthors}%
Liu, H.%
\BCBT {}\ \BBA {} L\"uhr, H.%
\end{APACrefauthors}%
\unskip\
\newblock
\APACrefYearMonthDay{2005}{}{}.
\newblock
{\BBOQ}\APACrefatitle {Strong disturbance of the upper thermospheric density
  due to magnetic storms: {CHAMP} observations} {Strong disturbance of the
  upper thermospheric density due to magnetic storms: {CHAMP}
  observations}.{\BBCQ}
\newblock
\APACjournalVolNumPages{Journal\ of\ Geophysical\ Research}{110}{A9}{1--9}.
\newblock
\begin{APACrefDOI} \doi{10.1029/2004JA010908} \end{APACrefDOI}
\PrintBackRefs{\CurrentBib}

\bibitem [\protect \citeauthoryear {%
Liu%
, L\"uhr%
, Henize%
\BCBL {}\ \BBA {} K\"ohler%
}{%
Liu%
\ \protect \BOthers {.}}{%
{\protect \APACyear {2005}}%
}]{%
Liu2005b}
\APACinsertmetastar {%
Liu2005b}%
\begin{APACrefauthors}%
Liu, H.%
, L\"uhr, H.%
, Henize, V.%
\BCBL {}\ \BBA {} K\"ohler, W.%
\end{APACrefauthors}%
\unskip\
\newblock
\APACrefYearMonthDay{2005}{}{}.
\newblock
{\BBOQ}\APACrefatitle {Global distribution of the thermospheric total mass
  density derived from {CHAMP}} {Global distribution of the thermospheric total
  mass density derived from {CHAMP}}.{\BBCQ}
\newblock
\APACjournalVolNumPages{Journal\ of\ Geophysical\ Research}{110}{A4}{}.
\newblock
\begin{APACrefDOI} \doi{10.1029/2004JA010741} \end{APACrefDOI}
\PrintBackRefs{\CurrentBib}

\bibitem [\protect \citeauthoryear {%
Lockyer%
}{%
Lockyer%
}{%
{\protect \APACyear {1903}}%
}]{%
Lockyer1903}
\APACinsertmetastar {%
Lockyer1903}%
\begin{APACrefauthors}%
Lockyer, W\BPBI J\BPBI S.%
\end{APACrefauthors}%
\unskip\
\newblock
\APACrefYearMonthDay{1903}{}{}.
\newblock
{\BBOQ}\APACrefatitle {{Magnetic Storms, Auror\AE{} and Solar Phenomena}}
  {{Magnetic Storms, Auror\AE{} and Solar Phenomena}}.{\BBCQ}
\newblock
\APACjournalVolNumPages{Nature}{69}{1775}{9-10}.
\newblock
\begin{APACrefDOI} \doi{10.1038/069009a0} \end{APACrefDOI}
\PrintBackRefs{\CurrentBib}

\bibitem [\protect \citeauthoryear {%
Love%
, Hayakawa%
\BCBL {}\ \BBA {} Cliver%
}{%
Love%
\ \protect \BOthers {.}}{%
{\protect \APACyear {2019}}%
{\protect \APACexlab {{\protect \BCnt {1}}}}}]{%
Love2019b}
\APACinsertmetastar {%
Love2019b}%
\begin{APACrefauthors}%
Love, J\BPBI J.%
, Hayakawa, H.%
\BCBL {}\ \BBA {} Cliver, E\BPBI W.%
\end{APACrefauthors}%
\unskip\
\newblock
\APACrefYearMonthDay{2019{\protect \BCnt {1}}}{}{}.
\newblock
{\BBOQ}\APACrefatitle {{Intensity and impact of the New York Railroad
  superstorm of May 1921}} {{Intensity and impact of the New York Railroad
  superstorm of May 1921}}.{\BBCQ}
\newblock
\APACjournalVolNumPages{Space\ Weather}{17}{8}{1281-1292}.
\newblock
\begin{APACrefDOI} \doi{10.1029/2019SW002250} \end{APACrefDOI}
\PrintBackRefs{\CurrentBib}

\bibitem [\protect \citeauthoryear {%
Love%
, Hayakawa%
\BCBL {}\ \BBA {} Cliver%
}{%
Love%
\ \protect \BOthers {.}}{%
{\protect \APACyear {2019}}%
{\protect \APACexlab {{\protect \BCnt {2}}}}}]{%
Love2019a}
\APACinsertmetastar {%
Love2019a}%
\begin{APACrefauthors}%
Love, J\BPBI J.%
, Hayakawa, H.%
\BCBL {}\ \BBA {} Cliver, E\BPBI W.%
\end{APACrefauthors}%
\unskip\
\newblock
\APACrefYearMonthDay{2019{\protect \BCnt {2}}}{}{}.
\newblock
{\BBOQ}\APACrefatitle {{On the Intensity of the Magnetic Superstorm of
  September 1909}} {{On the Intensity of the Magnetic Superstorm of September
  1909}}.{\BBCQ}
\newblock
\APACjournalVolNumPages{Space\ Weather}{17}{1}{37-45}.
\newblock
\begin{APACrefDOI} \doi{10.1029/2018SW002079} \end{APACrefDOI}
\PrintBackRefs{\CurrentBib}

\bibitem [\protect \citeauthoryear {%
Lu%
, Richmond%
, L\"{u}hr%
\BCBL {}\ \BBA {} Paxton%
}{%
Lu%
\ \protect \BOthers {.}}{%
{\protect \APACyear {2016}}%
}]{%
Lu2016a}
\APACinsertmetastar {%
Lu2016a}%
\begin{APACrefauthors}%
Lu, G.%
, Richmond, A\BPBI D.%
, L\"{u}hr, H.%
\BCBL {}\ \BBA {} Paxton, L.%
\end{APACrefauthors}%
\unskip\
\newblock
\APACrefYearMonthDay{2016}{}{}.
\newblock
{\BBOQ}\APACrefatitle {High-latitude energy input and its impact on the
  thermosphere} {High-latitude energy input and its impact on the
  thermosphere}.{\BBCQ}
\newblock
\APACjournalVolNumPages{Journal\ of\ Geophysical\ Research:\ Space\
  Physics}{121}{7}{7108--7124}.
\newblock
\begin{APACrefDOI} \doi{10.1002/2015JA022294} \end{APACrefDOI}
\PrintBackRefs{\CurrentBib}

\bibitem [\protect \citeauthoryear {%
Meng%
, Tsurutani%
\BCBL {}\ \BBA {} Mannucci%
}{%
Meng%
\ \protect \BOthers {.}}{%
{\protect \APACyear {2019}}%
}]{%
Meng2019}
\APACinsertmetastar {%
Meng2019}%
\begin{APACrefauthors}%
Meng, X.%
, Tsurutani, B\BPBI T.%
\BCBL {}\ \BBA {} Mannucci, A\BPBI J.%
\end{APACrefauthors}%
\unskip\
\newblock
\APACrefYearMonthDay{2019}{}{}.
\newblock
{\BBOQ}\APACrefatitle {{The Solar and Interplanetary Causes of Superstorms
  (Minimum Dst $\leq$ --250 nT) During the Space Age}} {{The Solar and
  Interplanetary Causes of Superstorms (Minimum Dst $\leq$ --250 nT) During the
  Space Age}}.{\BBCQ}
\newblock
\APACjournalVolNumPages{Journal\ of\ Geophysical\ Research:\ Space\
  Physics}{124}{6}{3926-3948}.
\newblock
\begin{APACrefDOI} \doi{10.1029/2018JA026425} \end{APACrefDOI}
\PrintBackRefs{\CurrentBib}

\bibitem [\protect \citeauthoryear {%
Mlynczak%
\ \protect \BOthers {.}}{%
Mlynczak%
\ \protect \BOthers {.}}{%
{\protect \APACyear {2003}}%
}]{%
Mlynczak2003}
\APACinsertmetastar {%
Mlynczak2003}%
\begin{APACrefauthors}%
Mlynczak, M\BPBI G.%
, {Martin-Torres}, F\BPBI J.%
, Russell, J.%
, Beaumont, K.%
, Jacobson, S.%
, Kozyra, J.%
, L\'opez-Puertas, M.%
, Funke, B.%
, Mertens, C.%
, Gordley, L.%
, Picard, R.%
, Winick, J.%
, Wintersteiner, P.%
\BCBL {}\ \BBA {} Paxton, L.%
\end{APACrefauthors}%
\unskip\
\newblock
\APACrefYearMonthDay{2003}{}{}.
\newblock
{\BBOQ}\APACrefatitle {{The natural thermostat of nitric oxide emission at 5.3
  $\mu$m in the thermosphere observed during the solar storms of April 2002}}
  {{The natural thermostat of nitric oxide emission at 5.3 $\mu$m in the
  thermosphere observed during the solar storms of April 2002}}.{\BBCQ}
\newblock
\APACjournalVolNumPages{Geophysical\ Research\ Letters}{30}{21}{}.
\newblock
\begin{APACrefDOI} \doi{10.1029/2003GL017693} \end{APACrefDOI}
\PrintBackRefs{\CurrentBib}

\bibitem [\protect \citeauthoryear {%
Moe%
\ \BBA {} Moe%
}{%
Moe%
\ \BBA {} Moe%
}{%
{\protect \APACyear {2005}}%
}]{%
Moe2005}
\APACinsertmetastar {%
Moe2005}%
\begin{APACrefauthors}%
Moe, K.%
\BCBT {}\ \BBA {} Moe, M\BPBI M.%
\end{APACrefauthors}%
\unskip\
\newblock
\APACrefYearMonthDay{2005}{}{}.
\newblock
{\BBOQ}\APACrefatitle {Gas-surface interactions and satellite drag
  coefficients} {Gas-surface interactions and satellite drag
  coefficients}.{\BBCQ}
\newblock
\APACjournalVolNumPages{Planetary\ and\ Space\ Science}{53}{8}{793-801}.
\newblock
\begin{APACrefDOI}\doi{10.1016/j.pss.2005.03.005}\end{APACrefDOI}
\PrintBackRefs{\CurrentBib}

\bibitem [\protect \citeauthoryear {%
{National Science and Technology Council}%
}{%
{National Science and Technology Council}%
}{%
{\protect \APACyear {2015}}%
{\protect \APACexlab {{\protect \BCnt {1}}}}}]{%
NSWAP2015}
\APACinsertmetastar {%
NSWAP2015}%
\begin{APACrefauthors}%
{National Science and Technology Council}.%
\end{APACrefauthors}%
\unskip\
\newblock
\APACrefYearMonthDay{2015{\protect \BCnt {1}}}{}{}.
\newblock
\APACrefbtitle {{National Space Weather Action Plan}} {{National Space Weather
  Action Plan}}\ \APACbVolEdTR{}{\BTR{}}.
\newblock
\APACaddressInstitution{Washington,\ D.C.}{{Executive Office of the President
  of the United States}}.
\newblock
\begin{APACrefURL}
  \url{https://obamawhitehouse.archives.gov/sites/default/files/microsites/ostp/final_nationalspaceweatheractionplan_20151028.pdf}
  \end{APACrefURL}
\PrintBackRefs{\CurrentBib}

\bibitem [\protect \citeauthoryear {%
{National Science and Technology Council}%
}{%
{National Science and Technology Council}%
}{%
{\protect \APACyear {2015}}%
{\protect \APACexlab {{\protect \BCnt {2}}}}}]{%
NSWS2015}
\APACinsertmetastar {%
NSWS2015}%
\begin{APACrefauthors}%
{National Science and Technology Council}.%
\end{APACrefauthors}%
\unskip\
\newblock
\APACrefYearMonthDay{2015{\protect \BCnt {2}}}{}{}.
\newblock
\APACrefbtitle {{National Space Weather Strategy}} {{National Space Weather
  Strategy}}\ \APACbVolEdTR{}{\BTR{}}.
\newblock
\APACaddressInstitution{Washington,\ D.C.}{{Executive Office of the President
  of the United States}}.
\newblock
\begin{APACrefURL}
  \url{https://obamawhitehouse.archives.gov/sites/default/files/microsites/ostp/final_nationalspaceweatherstrategy_20151028.pdf}
  \end{APACrefURL}
\PrintBackRefs{\CurrentBib}

\bibitem [\protect \citeauthoryear {%
Oliveira%
\ \protect \BOthers {.}}{%
Oliveira%
\ \protect \BOthers {.}}{%
{\protect \APACyear {2018}}%
}]{%
Oliveira2018b}
\APACinsertmetastar {%
Oliveira2018b}%
\begin{APACrefauthors}%
Oliveira, D\BPBI M.%
, Arel, D.%
, Raeder, J.%
, Zesta, E.%
, Ngwira, C\BPBI M.%
, Carter, B\BPBI A.%
, Yizengaw, E.%
, Halford, A\BPBI J.%
, Tsurutani, B\BPBI T.%
\BCBL {}\ \BBA {} Gjerloev, J\BPBI W.%
\end{APACrefauthors}%
\unskip\
\newblock
\APACrefYearMonthDay{2018}{}{}.
\newblock
{\BBOQ}\APACrefatitle {Geomagnetically induced currents caused by
  interplanetary shocks with different impact angles and speeds}
  {Geomagnetically induced currents caused by interplanetary shocks with
  different impact angles and speeds}.{\BBCQ}
\newblock
\APACjournalVolNumPages{Space\ Weather}{16}{6}{636-647}.
\newblock
\begin{APACrefDOI} \doi{10.1029/2018SW001880} \end{APACrefDOI}
\PrintBackRefs{\CurrentBib}

\bibitem [\protect \citeauthoryear {%
Oliveira%
\ \BBA {} Ngwira%
}{%
Oliveira%
\ \BBA {} Ngwira%
}{%
{\protect \APACyear {2017}}%
}]{%
Oliveira2017d}
\APACinsertmetastar {%
Oliveira2017d}%
\begin{APACrefauthors}%
Oliveira, D\BPBI M.%
\BCBT {}\ \BBA {} Ngwira, C\BPBI M.%
\end{APACrefauthors}%
\unskip\
\newblock
\APACrefYearMonthDay{2017}{}{}.
\newblock
{\BBOQ}\APACrefatitle {{Geomagnetically Induced Currents: Principles}}
  {{Geomagnetically Induced Currents: Principles}}.{\BBCQ}
\newblock
\APACjournalVolNumPages{Brazilian\ Journal\ of\ Physics}{47}{5}{552-560}.
\newblock
\begin{APACrefDOI} \doi{10.1007/s13538-017-0523-y} \end{APACrefDOI}
\PrintBackRefs{\CurrentBib}

\bibitem [\protect \citeauthoryear {%
Oliveira%
\ \BBA {} Zesta%
}{%
Oliveira%
\ \BBA {} Zesta%
}{%
{\protect \APACyear {2019}}%
}]{%
Oliveira2019b}
\APACinsertmetastar {%
Oliveira2019b}%
\begin{APACrefauthors}%
Oliveira, D\BPBI M.%
\BCBT {}\ \BBA {} Zesta, E.%
\end{APACrefauthors}%
\unskip\
\newblock
\APACrefYearMonthDay{2019}{}{}.
\newblock
{\BBOQ}\APACrefatitle {{Satellite Orbital Drag During Magnetic Storms}}
  {{Satellite Orbital Drag During Magnetic Storms}}.{\BBCQ}
\newblock
\APACjournalVolNumPages{Space\ Weather}{17}{11}{1510-1533}.
\newblock
\begin{APACrefDOI} \doi{10.1029/2019SW002287} \end{APACrefDOI}
\PrintBackRefs{\CurrentBib}

\bibitem [\protect \citeauthoryear {%
Oliveira%
, Zesta%
, Schuck%
\BCBL {}\ \BBA {} Sutton%
}{%
Oliveira%
\ \protect \BOthers {.}}{%
{\protect \APACyear {2017}}%
}]{%
Oliveira2017c}
\APACinsertmetastar {%
Oliveira2017c}%
\begin{APACrefauthors}%
Oliveira, D\BPBI M.%
, Zesta, E.%
, Schuck, P\BPBI W.%
\BCBL {}\ \BBA {} Sutton, E\BPBI K.%
\end{APACrefauthors}%
\unskip\
\newblock
\APACrefYearMonthDay{2017}{}{}.
\newblock
{\BBOQ}\APACrefatitle {Thermosphere global time response to geomagnetic storms
  caused by coronal mass ejections} {Thermosphere global time response to
  geomagnetic storms caused by coronal mass ejections}.{\BBCQ}
\newblock
\APACjournalVolNumPages{Journal\ of\ Geophysical\ Research:\ Space\
  Physics}{122}{10}{10,762-10,782}.
\newblock
\begin{APACrefDOI} \doi{10.1002/2017JA024006} \end{APACrefDOI}
\PrintBackRefs{\CurrentBib}

\bibitem [\protect \citeauthoryear {%
Page%
}{%
Page%
}{%
{\protect \APACyear {1903}}%
}]{%
Page1903}
\APACinsertmetastar {%
Page1903}%
\begin{APACrefauthors}%
Page, J.%
\end{APACrefauthors}%
\unskip\
\newblock
\APACrefYearMonthDay{1903}{}{}.
\newblock
{\BBOQ}\APACrefatitle {{The Polar Aurora of October 30 -- November 1, 1903}}
  {{The Polar Aurora of October 30 -- November 1, 1903}}.{\BBCQ}
\newblock
\APACjournalVolNumPages{{Monthly\ Weather\ Review}}{31}{12}{592-593}.
\newblock
\begin{APACrefDOI} \doi{10.1175/1520-0493(1903)31[592c:TPAOON]2.0.CO;2}
  \end{APACrefDOI}
\PrintBackRefs{\CurrentBib}

\bibitem [\protect \citeauthoryear {%
Picone%
, Hedin%
, Drob%
\BCBL {}\ \BBA {} Aikin%
}{%
Picone%
\ \protect \BOthers {.}}{%
{\protect \APACyear {2002}}%
}]{%
Picone2002}
\APACinsertmetastar {%
Picone2002}%
\begin{APACrefauthors}%
Picone, J\BPBI M.%
, Hedin, A\BPBI E.%
, Drob, D\BPBI P.%
\BCBL {}\ \BBA {} Aikin, A\BPBI C.%
\end{APACrefauthors}%
\unskip\
\newblock
\APACrefYearMonthDay{2002}{}{}.
\newblock
{\BBOQ}\APACrefatitle {{NRLMSISE-00} empirical model of the atmosphere:
  {S}tatistical comparisons and scientific issues} {{NRLMSISE-00} empirical
  model of the atmosphere: {S}tatistical comparisons and scientific
  issues}.{\BBCQ}
\newblock
\APACjournalVolNumPages{Journal\ of\ Geophysical\ Research}{107}{A12}{SIA
  15-1--SIA 15-16}.
\newblock
\begin{APACrefDOI} \doi{10.1029/2002JA009430} \end{APACrefDOI}
\PrintBackRefs{\CurrentBib}

\bibitem [\protect \citeauthoryear {%
Prieto%
, Graziano%
\BCBL {}\ \BBA {} Roberts%
}{%
Prieto%
\ \protect \BOthers {.}}{%
{\protect \APACyear {2014}}%
}]{%
Prieto2014}
\APACinsertmetastar {%
Prieto2014}%
\begin{APACrefauthors}%
Prieto, D\BPBI M.%
, Graziano, B\BPBI P.%
\BCBL {}\ \BBA {} Roberts, P\BPBI C\BPBI E.%
\end{APACrefauthors}%
\unskip\
\newblock
\APACrefYearMonthDay{2014}{}{}.
\newblock
{\BBOQ}\APACrefatitle {Spacecraft drag modelling} {Spacecraft drag
  modelling}.{\BBCQ}
\newblock
\APACjournalVolNumPages{Progress\ in\ Aerospace\ Sciences}{64}{}{56-65}.
\newblock
\begin{APACrefDOI} \doi{10.1016/j.paerosci.2013.09.001} \end{APACrefDOI}
\PrintBackRefs{\CurrentBib}

\bibitem [\protect \citeauthoryear {%
Pr\"olss%
}{%
Pr\"olss%
}{%
{\protect \APACyear {2011}}%
}]{%
Prolss2011}
\APACinsertmetastar {%
Prolss2011}%
\begin{APACrefauthors}%
Pr\"olss, G.%
\end{APACrefauthors}%
\unskip\
\newblock
\APACrefYearMonthDay{2011}{}{}.
\newblock
{\BBOQ}\APACrefatitle {Density Perturbations in the Upper Atmosphere Caused by
  the Dissipation of Solar Wind Energy} {Density perturbations in the upper
  atmosphere caused by the dissipation of solar wind energy}.{\BBCQ}
\newblock
\APACjournalVolNumPages{Surveys\ in\ Geophysics}{32}{2}{101--195}.
\newblock
\begin{APACrefDOI} \doi{10.1007/s10712-010-9104-0} \end{APACrefDOI}
\PrintBackRefs{\CurrentBib}

\bibitem [\protect \citeauthoryear {%
Pulkkinen%
, Bernabeu%
, Eichner%
, Beggan%
\BCBL {}\ \BBA {} Thomson%
}{%
Pulkkinen%
\ \protect \BOthers {.}}{%
{\protect \APACyear {2012}}%
}]{%
Pulkkinen2012}
\APACinsertmetastar {%
Pulkkinen2012}%
\begin{APACrefauthors}%
Pulkkinen, A.%
, Bernabeu, E.%
, Eichner, J.%
, Beggan, C.%
\BCBL {}\ \BBA {} Thomson, A\BPBI W\BPBI P.%
\end{APACrefauthors}%
\unskip\
\newblock
\APACrefYearMonthDay{2012}{}{}.
\newblock
{\BBOQ}\APACrefatitle {Generation of 100-year geomagnetically induced current
  scenarios} {Generation of 100-year geomagnetically induced current
  scenarios}.{\BBCQ}
\newblock
\APACjournalVolNumPages{Space\ Weather}{10}{4}{}.
\newblock
\begin{APACrefDOI} \doi{10.1029/2011SW000750} \end{APACrefDOI}
\PrintBackRefs{\CurrentBib}

\bibitem [\protect \citeauthoryear {%
Pulkkinen%
\ \protect \BOthers {.}}{%
Pulkkinen%
\ \protect \BOthers {.}}{%
{\protect \APACyear {2017}}%
}]{%
Pulkkinen2017}
\APACinsertmetastar {%
Pulkkinen2017}%
\begin{APACrefauthors}%
Pulkkinen, A.%
, Bernabeu, E.%
, Thomson, A.%
, Viljanen, A.%
, Pirjola, R.%
, Boteler, D.%
, Eichner, J.%
, Cilliers, P\BPBI J.%
, Welling, D.%
, Savani, N\BPBI P.%
, Weigel, R\BPBI S.%
, Love, J\BPBI J.%
, Balch, C.%
, Ngwira, C\BPBI M.%
, Crowley, G.%
, Schultz, A.%
, Kataoka, R.%
, Anderson, B.%
, Fugate, D.%
, Simpson, J\BPBI J.%
\BCBL {}\ \BBA {} MacAlester, M.%
\end{APACrefauthors}%
\unskip\
\newblock
\APACrefYearMonthDay{2017}{}{}.
\newblock
{\BBOQ}\APACrefatitle {{Geomagnetically induced currents: Science, engineering,
  and applications readiness}} {{Geomagnetically induced currents: Science,
  engineering, and applications readiness}}.{\BBCQ}
\newblock
\APACjournalVolNumPages{Space\ Weather}{15}{7}{828-856}.
\newblock
\begin{APACrefDOI} \doi{10.1002/2016SW001501} \end{APACrefDOI}
\PrintBackRefs{\CurrentBib}

\bibitem [\protect \citeauthoryear {%
Reigber%
, L\"uhr%
\BCBL {}\ \BBA {} Schwintzer%
}{%
Reigber%
\ \protect \BOthers {.}}{%
{\protect \APACyear {2002}}%
}]{%
Reigber2002a}
\APACinsertmetastar {%
Reigber2002a}%
\begin{APACrefauthors}%
Reigber, C.%
, L\"uhr, H.%
\BCBL {}\ \BBA {} Schwintzer, P.%
\end{APACrefauthors}%
\unskip\
\newblock
\APACrefYearMonthDay{2002}{}{}.
\newblock
{\BBOQ}\APACrefatitle {{CHAMP} mission status} {{CHAMP} mission status}.{\BBCQ}
\newblock
\APACjournalVolNumPages{Advances\ in\ Space\ {Research}}{30}{2}{129-134}.
\newblock
\begin{APACrefDOI} \doi{10.1016/S0273-1177(02)00276-4} \end{APACrefDOI}
\PrintBackRefs{\CurrentBib}

\bibitem [\protect \citeauthoryear {%
Ribeiro%
, Vaquero%
, Gallego%
\BCBL {}\ \BBA {} Trigo%
}{%
Ribeiro%
\ \protect \BOthers {.}}{%
{\protect \APACyear {2016}}%
}]{%
Ribeiro2016}
\APACinsertmetastar {%
Ribeiro2016}%
\begin{APACrefauthors}%
Ribeiro, P.%
, Vaquero, J\BPBI M.%
, Gallego, M\BPBI C.%
\BCBL {}\ \BBA {} Trigo, R\BPBI M.%
\end{APACrefauthors}%
\unskip\
\newblock
\APACrefYearMonthDay{2016}{}{}.
\newblock
{\BBOQ}\APACrefatitle {{The First Documented Space Weather Event That Perturbed
  the Communication Networks in Iberia}} {{The First Documented Space Weather
  Event That Perturbed the Communication Networks in Iberia}}.{\BBCQ}
\newblock
\APACjournalVolNumPages{Space\ Weather}{14}{7}{464-468}.
\newblock
\begin{APACrefDOI} \doi{10.1002/2016SW001424} \end{APACrefDOI}
\PrintBackRefs{\CurrentBib}

\bibitem [\protect \citeauthoryear {%
Rich%
\ \BBA {} Denig%
}{%
Rich%
\ \BBA {} Denig%
}{%
{\protect \APACyear {1992}}%
}]{%
Rich1992}
\APACinsertmetastar {%
Rich1992}%
\begin{APACrefauthors}%
Rich, F\BPBI J.%
\BCBT {}\ \BBA {} Denig, W\BPBI F.%
\end{APACrefauthors}%
\unskip\
\newblock
\APACrefYearMonthDay{1992}{}{}.
\newblock
{\BBOQ}\APACrefatitle {{The major magnetic storm of March 13-14, 1989 and
  associated ionosphere effects}} {{The major magnetic storm of March 13-14,
  1989 and associated ionosphere effects}}.{\BBCQ}
\newblock
\APACjournalVolNumPages{Canadian\ Journal\ of\ Physics}{70}{7}{510-525}.
\newblock
\begin{APACrefDOI} \doi{10.1139/p92-086} \end{APACrefDOI}
\PrintBackRefs{\CurrentBib}

\bibitem [\protect \citeauthoryear {%
Richmond%
\ \BBA {} Lu%
}{%
Richmond%
\ \BBA {} Lu%
}{%
{\protect \APACyear {2000}}%
}]{%
Richmond2000}
\APACinsertmetastar {%
Richmond2000}%
\begin{APACrefauthors}%
Richmond, A.%
\BCBT {}\ \BBA {} Lu, G.%
\end{APACrefauthors}%
\unskip\
\newblock
\APACrefYearMonthDay{2000}{}{}.
\newblock
{\BBOQ}\APACrefatitle {{Upper-atmospheric effects of magnetic storms: A brief
  tutorial}} {{Upper-atmospheric effects of magnetic storms: A brief
  tutorial}}.{\BBCQ}
\newblock
\APACjournalVolNumPages{Journal\ of\ Atmospheric\ and\ Solar-Terrestrial\
  Physics}{62}{12}{1115-1127}.
\newblock
\begin{APACrefDOI} \doi{10.1016/S1364-6826(00)00094-8} \end{APACrefDOI}
\PrintBackRefs{\CurrentBib}

\bibitem [\protect \citeauthoryear {%
Riley%
\ \protect \BOthers {.}}{%
Riley%
\ \protect \BOthers {.}}{%
{\protect \APACyear {2018}}%
}]{%
Riley2018}
\APACinsertmetastar {%
Riley2018}%
\begin{APACrefauthors}%
Riley, P.%
, Baker, D.%
, Liu, Y\BPBI D.%
, Verronen, P.%
, Singer, H.%
\BCBL {}\ \BBA {} G\"udel, M.%
\end{APACrefauthors}%
\unskip\
\newblock
\APACrefYearMonthDay{2018}{}{}.
\newblock
{\BBOQ}\APACrefatitle {{Extreme Space Weather Events: From Cradle to Grave}}
  {{Extreme Space Weather Events: From Cradle to Grave}}.{\BBCQ}
\newblock
\APACjournalVolNumPages{Space\ Science\ Reviews}{214}{21}{}.
\newblock
\begin{APACrefDOI} \doi{10.1007/s11214-017-0456-3} \end{APACrefDOI}
\PrintBackRefs{\CurrentBib}

\bibitem [\protect \citeauthoryear {%
Shepherd%
}{%
Shepherd%
}{%
{\protect \APACyear {2014}}%
}]{%
Shepherd2014}
\APACinsertmetastar {%
Shepherd2014}%
\begin{APACrefauthors}%
Shepherd, S\BPBI G.%
\end{APACrefauthors}%
\unskip\
\newblock
\APACrefYearMonthDay{2014}{}{}.
\newblock
{\BBOQ}\APACrefatitle {Altitude-adjusted corrected geomagnetic coordinates:
  {D}efinition and functional approximations} {Altitude-adjusted corrected
  geomagnetic coordinates: {D}efinition and functional approximations}.{\BBCQ}
\newblock
\APACjournalVolNumPages{Journal\ of\ Geophysical\ Research:\ Space\
  Physics}{119}{9}{7501--7521}.
\newblock
\begin{APACrefDOI} \doi{10.1002/2014JA020264} \end{APACrefDOI}
\PrintBackRefs{\CurrentBib}

\bibitem [\protect \citeauthoryear {%
Shi%
\ \protect \BOthers {.}}{%
Shi%
\ \protect \BOthers {.}}{%
{\protect \APACyear {2019}}%
}]{%
Shi2019b}
\APACinsertmetastar {%
Shi2019b}%
\begin{APACrefauthors}%
Shi, Y.%
, Oliveira, D\BPBI M.%
, Knipp, D\BPBI J.%
, Zesta, E.%
, Matsuo, T.%
\BCBL {}\ \BBA {} Anderson, B.%
\end{APACrefauthors}%
\unskip\
\newblock
\APACrefYearMonthDay{2019}{}{}.
\newblock
{\BBOQ}\APACrefatitle {{Effects of Nearly Frontal and Highly Inclined
  Interplanetary Shocks on High-latitude Field-aligned Currents (FACs)}}
  {{Effects of Nearly Frontal and Highly Inclined Interplanetary Shocks on
  High-latitude Field-aligned Currents (FACs)}}.{\BBCQ}
\newblock
\APACjournalVolNumPages{Space\ Weather}{17}{12}{1659-1673}.
\newblock
\begin{APACrefDOI} \doi{10.1029/2019SW002367} \end{APACrefDOI}
\PrintBackRefs{\CurrentBib}

\bibitem [\protect \citeauthoryear {%
Silverman%
}{%
Silverman%
}{%
{\protect \APACyear {1995}}%
}]{%
Silverman1995}
\APACinsertmetastar {%
Silverman1995}%
\begin{APACrefauthors}%
Silverman, S\BPBI M.%
\end{APACrefauthors}%
\unskip\
\newblock
\APACrefYearMonthDay{1995}{}{}.
\newblock
{\BBOQ}\APACrefatitle {Low latitude auroras: the storm of 25 {September} 1909}
  {Low latitude auroras: the storm of 25 {September} 1909}.{\BBCQ}
\newblock
\APACjournalVolNumPages{Journal\ of\ Atmospheric\ and\ Solar-Terrestrial\
  Physics}{57}{6}{673-685}.
\newblock
\begin{APACrefDOI} \doi{10.1016/0021-9169(94)E0012-C} \end{APACrefDOI}
\PrintBackRefs{\CurrentBib}

\bibitem [\protect \citeauthoryear {%
Silverman%
\ \BBA {} Cliver%
}{%
Silverman%
\ \BBA {} Cliver%
}{%
{\protect \APACyear {2001}}%
}]{%
Silverman2001}
\APACinsertmetastar {%
Silverman2001}%
\begin{APACrefauthors}%
Silverman, S\BPBI M.%
\BCBT {}\ \BBA {} Cliver, E\BPBI W.%
\end{APACrefauthors}%
\unskip\
\newblock
\APACrefYearMonthDay{2001}{}{}.
\newblock
{\BBOQ}\APACrefatitle {{Low-latitude auroras: the magnetic storm of 14-15 May
  1921}} {{Low-latitude auroras: the magnetic storm of 14-15 May 1921}}.{\BBCQ}
\newblock
\APACjournalVolNumPages{Journal\ of\ Atmospheric\ and\ Solar-Terrestrial\
  Physics}{63}{5}{523-535}.
\newblock
\begin{APACrefDOI} \doi{10.1016/S1364-6826(00)00174-7} \end{APACrefDOI}
\PrintBackRefs{\CurrentBib}

\bibitem [\protect \citeauthoryear {%
Siscoe%
, Crooker%
\BCBL {}\ \BBA {} Clauer%
}{%
Siscoe%
\ \protect \BOthers {.}}{%
{\protect \APACyear {2006}}%
}]{%
Siscoe2006}
\APACinsertmetastar {%
Siscoe2006}%
\begin{APACrefauthors}%
Siscoe, G.%
, Crooker, N\BPBI U.%
\BCBL {}\ \BBA {} Clauer, C\BPBI R.%
\end{APACrefauthors}%
\unskip\
\newblock
\APACrefYearMonthDay{2006}{}{}.
\newblock
{\BBOQ}\APACrefatitle {{Dst of the Carrington storm of 1859}} {{Dst of the
  Carrington storm of 1859}}.{\BBCQ}
\newblock
\APACjournalVolNumPages{Advances\ in\ Space\ {Research}}{38}{2}{173-179}.
\newblock
\begin{APACrefDOI}\doi{10.1016/j.asr.2005.02.102}\end{APACrefDOI}
\PrintBackRefs{\CurrentBib}

\bibitem [\protect \citeauthoryear {%
Storz%
, Bowman%
, Branson%
, Casalic%
\BCBL {}\ \BBA {} Tobiska%
}{%
Storz%
\ \protect \BOthers {.}}{%
{\protect \APACyear {2005}}%
}]{%
Storz2005}
\APACinsertmetastar {%
Storz2005}%
\begin{APACrefauthors}%
Storz, M\BPBI F.%
, Bowman, B\BPBI R.%
, Branson, J\BPBI I.%
, Casalic, S\BPBI J.%
\BCBL {}\ \BBA {} Tobiska, W\BPBI K.%
\end{APACrefauthors}%
\unskip\
\newblock
\APACrefYearMonthDay{2005}{}{}.
\newblock
{\BBOQ}\APACrefatitle {{High accuracy satellite drag model (HASDM)}} {{High
  accuracy satellite drag model (HASDM)}}.{\BBCQ}
\newblock
\APACjournalVolNumPages{Advances\ in\ Space\ {Research}}{36}{12}{2497-2505}.
\newblock
\begin{APACrefDOI} \doi{10.1016/j.asr.2004.02.020} \end{APACrefDOI}
\PrintBackRefs{\CurrentBib}

\bibitem [\protect \citeauthoryear {%
Sugiura%
}{%
Sugiura%
}{%
{\protect \APACyear {1964}}%
}]{%
Sugiura1964a}
\APACinsertmetastar {%
Sugiura1964a}%
\begin{APACrefauthors}%
Sugiura, M.%
\end{APACrefauthors}%
\unskip\
\newblock
\APACrefYearMonthDay{1964}{}{}.
\newblock
{\BBOQ}\APACrefatitle {Hourly values of equatorial {Dst} for the {IGY}} {Hourly
  values of equatorial {Dst} for the {IGY}}.{\BBCQ}
\newblock
\APACjournalVolNumPages{Ann.\ Int.\ Geophys.\ Year}{35}{5}{}.
\PrintBackRefs{\CurrentBib}

\bibitem [\protect \citeauthoryear {%
Sutton%
}{%
Sutton%
}{%
{\protect \APACyear {2009}}%
}]{%
Sutton2009b}
\APACinsertmetastar {%
Sutton2009b}%
\begin{APACrefauthors}%
Sutton, E\BPBI K.%
\end{APACrefauthors}%
\unskip\
\newblock
\APACrefYearMonthDay{2009}{}{}.
\newblock
{\BBOQ}\APACrefatitle {Normalized force coefficients for satellites with
  elongated shapes} {Normalized force coefficients for satellites with
  elongated shapes}.{\BBCQ}
\newblock
\APACjournalVolNumPages{Journal\ of\ Spacecraft\ and\
  Rockets}{46}{1}{112--116}.
\newblock
\begin{APACrefDOI}\doi{10.2514/1.40940}\end{APACrefDOI}
\PrintBackRefs{\CurrentBib}

\bibitem [\protect \citeauthoryear {%
Sutton%
, Forbes%
\BCBL {}\ \BBA {} Knipp%
}{%
Sutton%
\ \protect \BOthers {.}}{%
{\protect \APACyear {2009}}%
}]{%
Sutton2009a}
\APACinsertmetastar {%
Sutton2009a}%
\begin{APACrefauthors}%
Sutton, E\BPBI K.%
, Forbes, J\BPBI M.%
\BCBL {}\ \BBA {} Knipp, D\BPBI J.%
\end{APACrefauthors}%
\unskip\
\newblock
\APACrefYearMonthDay{2009}{}{}.
\newblock
{\BBOQ}\APACrefatitle {Rapid response of the thermosphere to variations in
  {J}oule heating} {Rapid response of the thermosphere to variations in {J}oule
  heating}.{\BBCQ}
\newblock
\APACjournalVolNumPages{Journal\ of\ Geophysical\ Research}{114}{A4}{}.
\newblock
\begin{APACrefDOI} \doi{10.1029/2008JA013667} \end{APACrefDOI}
\PrintBackRefs{\CurrentBib}

\bibitem [\protect \citeauthoryear {%
Tapley%
, Bettadpur%
, Watkins%
\BCBL {}\ \BBA {} Reigber%
}{%
Tapley%
\ \protect \BOthers {.}}{%
{\protect \APACyear {2004}}%
}]{%
Tapley2004a}
\APACinsertmetastar {%
Tapley2004a}%
\begin{APACrefauthors}%
Tapley, B\BPBI D.%
, Bettadpur, S.%
, Watkins, M.%
\BCBL {}\ \BBA {} Reigber, C.%
\end{APACrefauthors}%
\unskip\
\newblock
\APACrefYearMonthDay{2004}{}{}.
\newblock
{\BBOQ}\APACrefatitle {The gravity recovery and climate experiment: Mission
  overview and early results} {The gravity recovery and climate experiment:
  Mission overview and early results}.{\BBCQ}
\newblock
\APACjournalVolNumPages{Geophysical\ Research\ Letters}{31}{9}{}.
\newblock
\begin{APACrefDOI} \doi{10.1029/2004GL019920} \end{APACrefDOI}
\PrintBackRefs{\CurrentBib}

\bibitem [\protect \citeauthoryear {%
Tsurutani%
, Gonzalez%
, Lakhina%
\BCBL {}\ \BBA {} Alex%
}{%
Tsurutani%
\ \protect \BOthers {.}}{%
{\protect \APACyear {2003}}%
}]{%
Tsurutani2003b}
\APACinsertmetastar {%
Tsurutani2003b}%
\begin{APACrefauthors}%
Tsurutani, B\BPBI T.%
, Gonzalez, W\BPBI D.%
, Lakhina, G\BPBI S.%
\BCBL {}\ \BBA {} Alex, S.%
\end{APACrefauthors}%
\unskip\
\newblock
\APACrefYearMonthDay{2003}{}{}.
\newblock
{\BBOQ}\APACrefatitle {{The extreme magnetic storm of 1–-2 September 1859}}
  {{The extreme magnetic storm of 1–-2 September 1859}}.{\BBCQ}
\newblock
\APACjournalVolNumPages{Journal\ of\ Geophysical\ Research}{108}{A7}{}.
\newblock
\begin{APACrefDOI} \doi{10.1029/2002JA009504} \end{APACrefDOI}
\PrintBackRefs{\CurrentBib}

\bibitem [\protect \citeauthoryear {%
Tsurutani%
\ \BBA {} Lakhina%
}{%
Tsurutani%
\ \BBA {} Lakhina%
}{%
{\protect \APACyear {2014}}%
}]{%
Tsurutani2014a}
\APACinsertmetastar {%
Tsurutani2014a}%
\begin{APACrefauthors}%
Tsurutani, B\BPBI T.%
\BCBT {}\ \BBA {} Lakhina, G\BPBI S.%
\end{APACrefauthors}%
\unskip\
\newblock
\APACrefYearMonthDay{2014}{}{}.
\newblock
{\BBOQ}\APACrefatitle {An extreme coronal mass ejection and consequences for
  the magnetosphere and {Earth}} {An extreme coronal mass ejection and
  consequences for the magnetosphere and {Earth}}.{\BBCQ}
\newblock
\APACjournalVolNumPages{Geophysical\ Research\ Letters}{41}{2}{287-292}.
\newblock
\begin{APACrefDOI} \doi{10.1002/2013GL058825} \end{APACrefDOI}
\PrintBackRefs{\CurrentBib}

\bibitem [\protect \citeauthoryear {%
Tsurutani%
\ \protect \BOthers {.}}{%
Tsurutani%
\ \protect \BOthers {.}}{%
{\protect \APACyear {2007}}%
}]{%
Tsurutani2007}
\APACinsertmetastar {%
Tsurutani2007}%
\begin{APACrefauthors}%
Tsurutani, B\BPBI T.%
, Verkhoglyadova, O\BPBI P.%
, Mannucci, A\BPBI J.%
, Araki, T.%
, Sato, A.%
, Tsuda, T.%
\BCBL {}\ \BBA {} Yumoto, K.%
\end{APACrefauthors}%
\unskip\
\newblock
\APACrefYearMonthDay{2007}{}{}.
\newblock
{\BBOQ}\APACrefatitle {{Oxygen ion up-lift and satellite drag effects during
  the 30 October 2003 daytime superfountain event}} {{Oxygen ion up-lift and
  satellite drag effects during the 30 October 2003 daytime superfountain
  event}}.{\BBCQ}
\newblock
\APACjournalVolNumPages{Annales\ Geophysicae}{25}{}{569–574}.
\newblock
\begin{APACrefDOI} \doi{10.5194/angeo-25-569-2007} \end{APACrefDOI}
\PrintBackRefs{\CurrentBib}

\bibitem [\protect \citeauthoryear {%
Vennerstr{\o}m%
\ \protect \BOthers {.}}{%
Vennerstr{\o}m%
\ \protect \BOthers {.}}{%
{\protect \APACyear {2016}}%
}]{%
Vennerstrom2016}
\APACinsertmetastar {%
Vennerstrom2016}%
\begin{APACrefauthors}%
Vennerstr{\o}m, S.%
, Lef\'evre, L.%
, Dumbovi\'c, M.%
, Crosby, N.%
, Malandraki, O.%
, Patsou, I.%
, Clette, F.%
, Veronig, A.%
, Vr\u{s}nak, B.%
, Leer, K.%
\BCBL {}\ \BBA {} Moretto, T.%
\end{APACrefauthors}%
\unskip\
\newblock
\APACrefYearMonthDay{2016}{}{}.
\newblock
{\BBOQ}\APACrefatitle {Extreme geomagnetix storms - 1868 - 2010} {Extreme
  geomagnetix storms - 1868 - 2010}.{\BBCQ}
\newblock
\APACjournalVolNumPages{Solar\ Physics}{291}{5}{1447-1481}.
\newblock
\begin{APACrefDOI} \doi{10.1007/s11207-016-0897-y} \end{APACrefDOI}
\PrintBackRefs{\CurrentBib}

\bibitem [\protect \citeauthoryear {%
{World Data Center for Geomagnetism, Kyoto}%
, Nose%
, Iyemori%
, Sugiura%
\BCBL {}\ \BBA {} Kamei%
}{%
{World Data Center for Geomagnetism, Kyoto}%
\ \protect \BOthers {.}}{%
{\protect \APACyear {2015}}%
}]{%
WDC_Dst2015}
\APACinsertmetastar {%
WDC_Dst2015}%
\begin{APACrefauthors}%
{World Data Center for Geomagnetism, Kyoto}%
, Nose, M.%
, Iyemori, T.%
, Sugiura, M.%
\BCBL {}\ \BBA {} Kamei, T.%
\end{APACrefauthors}%
\unskip\
\newblock
\APACrefYearMonthDay{2015}{}{}.
\newblock
\APACrefbtitle {{Geomagnetic Dst index}.} {{Geomagnetic Dst index}.}
\newblock
\begin{APACrefDOI} \doi{10.17593/14515-74000} \end{APACrefDOI}
\PrintBackRefs{\CurrentBib}

\bibitem [\protect \citeauthoryear {%
Zesta%
\ \BBA {} Huang%
}{%
Zesta%
\ \BBA {} Huang%
}{%
{\protect \APACyear {2016}}%
}]{%
Zesta2016b}
\APACinsertmetastar {%
Zesta2016b}%
\begin{APACrefauthors}%
Zesta, E.%
\BCBT {}\ \BBA {} Huang, C\BPBI Y.%
\end{APACrefauthors}%
\unskip\
\newblock
\APACrefYearMonthDay{2016}{}{}.
\newblock
{\BBOQ}\APACrefatitle {Satellite orbital drag} {Satellite orbital drag}.{\BBCQ}
\newblock
\BIn{} G\BPBI V.~Khazanov\ (\BED), \APACrefbtitle {{Space Weather
  Fundamentals}} {{Space Weather Fundamentals}}\ (\BPGS\ 329--351).
\newblock
\APACaddressPublisher{Boca Raton, FL}{CRC Press}.
\PrintBackRefs{\CurrentBib}

\bibitem [\protect \citeauthoryear {%
Zesta%
\ \BBA {} Oliveira%
}{%
Zesta%
\ \BBA {} Oliveira%
}{%
{\protect \APACyear {2019}}%
}]{%
Zesta2019a}
\APACinsertmetastar {%
Zesta2019a}%
\begin{APACrefauthors}%
Zesta, E.%
\BCBT {}\ \BBA {} Oliveira, D\BPBI M.%
\end{APACrefauthors}%
\unskip\
\newblock
\APACrefYearMonthDay{2019}{}{}.
\newblock
{\BBOQ}\APACrefatitle {Thermospheric heating and cooling times during
  geomagnetic storms, including extreme events} {Thermospheric heating and
  cooling times during geomagnetic storms, including extreme events}.{\BBCQ}
\newblock
\APACjournalVolNumPages{Geophysical\ Research\ Letters}{46}{22}{12,739-12,746}.
\newblock
\begin{APACrefDOI} \doi{10.1029/2019GL085120} \end{APACrefDOI}
\PrintBackRefs{\CurrentBib}

\bibitem [\protect \citeauthoryear {%
Zhao%
\ \BBA {} Dryer%
}{%
Zhao%
\ \BBA {} Dryer%
}{%
{\protect \APACyear {2014}}%
}]{%
Zhao2014b}
\APACinsertmetastar {%
Zhao2014b}%
\begin{APACrefauthors}%
Zhao, X.%
\BCBT {}\ \BBA {} Dryer, M.%
\end{APACrefauthors}%
\unskip\
\newblock
\APACrefYearMonthDay{2014}{}{}.
\newblock
{\BBOQ}\APACrefatitle {Current status of {CME/shock} arrival time prediction}
  {Current status of {CME/shock} arrival time prediction}.{\BBCQ}
\newblock
\APACjournalVolNumPages{Space\ Weather}{12}{7}{448-469}.
\newblock
\begin{APACrefDOI} \doi{10.1002/2014SW001060} \end{APACrefDOI}
\PrintBackRefs{\CurrentBib}

\end{thebibliography}

\end{document}